\documentclass[12pt,preprint]{aastex}
\usepackage{lscape}
\usepackage{textcomp}
\usepackage{graphicx}
\usepackage{epsfig}
\usepackage{txfonts}
\shorttitle{Chemical properties of N$_2$H$^+$} \shortauthors{Naiping Yu
et al.}

\begin{document}


\title {Chemical evolution of N$_2$H$^+$ in six massive star-forming regions }
\author
 {Nai-Ping Yu\altaffilmark{1},Jin-Long Xu\altaffilmark{1},Jun-Jie Wang\altaffilmark{1},Xiao-Lan Liu\altaffilmark{1}}

\altaffiltext{1} {National Astronomical Observatories, Chinese
Academy of Sciences, Beijing 100012, China}

%\date{Accepted 1993 December 11. Received 1993 March 17}
%\pagerange{\pageref{firstpage}--\pageref{lastpage}} \pubyear{2004}

\label{firstpage}


\begin{abstract}
To investigate how the abundance of N$_2$H$^+$ varies as massive clumps evolve, here we present a multi-wavelength study toward six molecular clouds. All of these clouds contain several massive clumps in different evolutionary stages of star formation. Using archival data of Herschel InfraRed Galactic Plane Survey (Hi-GAL), we made H$_2$ column density and dust temperature maps of these regions by the spectral energy distribution (SED) method. We found all of the six clouds show distinct dust temperature gradients, ranging from $\sim$ 20 K to $\sim$ 30 K. This makes them good candidates to study chemical evolution of molecules (such as N$_2$H$^+$) in different evolutionary stages of star formation. Our molecular line data come from Millimeter Astronomy Legacy Team Survey at 90 GHz (MALT90). We made column density and then abundance maps of N$_2$H$^+$. We found that when the dust temperature is above 27 K, the abundance of N$_2$H$^+$ begins to decrease or reaches a plateau. We regard this is because that in the photodissociation regions (PDRs) around classical HII regions, N$_2$H$^+$ is destroyed by free electrons heavily. However, when the dust temperature is below 27 K, the abundance of N$_2$H$^+$ increases with dust temperature. This seems to be inconsistent with previous chemical models made in low-mass star-forming regions. In order to check out whether this inconsistency is caused by a different chemistry in high-mass star-forming clumps, higher angular resolution observations are necessary.
\end{abstract}


%_____________________________________________________________________

\section{Introduction}
Massive stars ($\geq$ 8 M$_\odot$) play an important role in the evolution of galaxies and molecular clouds. They release large amounts of energy into their surrounding interstellar medium (ISM) and have an immense impact on the subsequent star formation therein. They enrich heavy elements of the cosmic space in the form of supernova explosions. Massive stars are rare and used to form in dense molecular clouds at far distances from the solar system. This makes us hard to make a clear picture of the processes of a massive star formation. In the last few decades, both in theories and observations, a lot of research have been done to understand the formation of massive stars (e.g. Zinnecher \& Yorke 2007; Deharveng et al. 2010). Generally speaking, high-mass stars evolve through prestellar cores in starless cores to protostars in hot cores, and then to hyper-compact HII region (HCHII) or ultra-compact HII region (UCHII) when substantial UV photons and ionized stellar winds rapidly ionize the surrounding hydrogen. The final stages are compact and classical HII regions. However, compared with low-mass stars, the formation of a high-mass star is still not well understood. It is regarded that the chemical properties of molecular clouds undergoing star formation should also be different due to the physical changes in different star formation processes (e.g. Sakai et al. 2008; Sanhueza et al. 2012; Vasyunina et al. 2011). Although many studies have been done to understand massive star formation, less is studied about their chemistry. What is the chemistry of massive star formations? Are they different from their low-mass counterparts? Can the chemical evolution in massive star-forming regions be used as a chemical clock? Recently, several studies have focused on this subject (e.g. Vasyunina et al. 2011; Hoq et al. 2013; Miettinen 2014; Yu $\&$ Wang 2015; Yu $\&$ Xu 2016 ).

Compared with CO species, N$_2$H$^+$ is more resistant to freeze-out onto grains, thus it is regarded as a good tracer of dense gas in the early stages of star formation (Bergin et al. 2001). CO would easily be depleted onto the dust grains when dust temperature is below 20 K. Lee et al. (2004) combined a sequence of Bonnor-Ebert spheres and the inside-out collapse model to describe dynamics from the pre-protostellar stage to later stages. They found N$_2$H$^+$ is primarily formed through the gas-phase reaction H$_3$$^+$ + N$_2$ $\rightarrow$ N$_2$H$^+$ + H$_2$, and destroyed by CO molecules in the gas phase. So it should be expected that in the early stages of star formation, the abundance of N$_2$H$^+$ is relatively high as the depletion of CO in gas phase. As the central protostar evolves, the gas gets warm, and CO molecules begin to evaporate from the dust grains when dust temperature is above 20 K (Tobin et al. 2013). CO could destroy N$_2$H$^+$ through the reaction N$_2$H$^+$ + CO $\rightarrow$ HCO$^+$ + N$_2$ ( e.g. J${\o}$rgensen et al. 2004; Lee et al. 2004). According to the work of Vigren et al. (2012), N$_2$H$^+$ can also be destroyed by free electrons in HII regions: N$_2$H$^+$ + e$^-$ $\rightarrow$ N$_2$ + H or NH + N. Some researches have been done to check the chemical evolution of N$_2$H$^+$ in massive star-forming regions. Using the molecular line maps from Year 1 of the MALT90 Survey, Hoq et al. (2013) found the abundance of N$_2$H$^+$ increase as a function of evolutionary stage. Sanhueza et al. (2012) also found both the column density and abundance of N$_2$H$^+$ increase as the clumps evolve from ``quiescent" (starless candidates) to ``red" stages (HII region candidates). These results are inconsistent with predictions of chemical models introduced above. As these authors analyze, several reasons may account this phenomena: (i) Chemical models of low-mass star forming regions (e.g. Bergin $\&$ Langer 1997; Lee et al. 2004) focused on single low-mass cores, which cannot be compared with a clump (hosting tens or hundreds of cores). The Mopra beam is 38$^\prime$$^\prime$, this makes it impossible to resolve a single massive star-forming core. The Mopra telescope not only probes the warm core gas but also the surrounding cold diffuse material; (ii)The chemistry of N$_2$H$^+$ in massive star-forming regions may be really different from their low-mass counterparts; (iii)The formation and destruction rate of N$_2$H$^+$ might not be as accurate as is currently believed. We should also mention here that the initial conditions may be very different in different molecular clouds. For example, in the Galaxy, the $^{12}$C/$^{13}$C ratio ranges from $\sim$ 20 to $\sim$ 70, depending on the distance to the Galactic center (e.g. Savage et al. 2002). This might make their results not statistically significant. Here we present multi-wavelength study toward six massive star-forming regions containing several clumps in different evolutionary stages of star formation, with the aim to make a more clear picture of chemical evolution of N$_2$H$^+$ in massive star-forming regions. With only six molecular clouds, our work cannot be statistically significant, but given that the clumps in each cloud likely have similar initial conditions, we wish to use a different approach to found out whether our result will be consistent with previous work. We introduce our sources and data in Section 2, analysis is given in Section 3, results and discussions are in Section 4, and finally we summarize in Section 5.

\section{Source and Data}
In the followings we present our source introduction in Section 2.1, molecular data of MALT90 in Section 2.2, and far-infrared data from ATLASGAL and Hi-GAL in Section 2.3.
\subsection{Source selection}
 The source sample of this paper involves two filamentary clouds identified by Li et al. (2016), two bubbles from Churchwell et al. (2006; 2007), and two dense clouds from Rathborne et al. (2016). The basic information of them is shown in table 1. We can see that each source involves at least four dense clumps from the catalogue of Contreras et al. (2013). Guzm\'{a}n et al. (2015) classified the clumps into ``Quiescent", ``Protostellar", ``HII region" and ``PDRs" stages according to the schematic timeline of a massive star formation. All of the six sources involves at least two stages. The effective radius of most clumps are larger than 38$^\prime$$^\prime$, which means they can be resolved by the Mopra telescope. For G351.776-0.527, CN148 and S36, previous studies have already indicated they are candidates of massive star-forming regions (e.g. Klaassen et al. 2015; Dewangan et al. 2015; Torii et al. 2017). For the other three sources, all of them involves at least one clumps with masses $>$ 10$^3$ M$_\odot$. Given a typical star formation efficiency of 10$\%$-30$\%$ (Lada et al. 2010) and a cluster having a Salpeter-type initial stellar mass function (IMF), we could expect a 10$^3$ M$_\odot$ clump to form a star cluster with massive stars $>$ 20 M$_\odot$. Therefore, our sources are candidates of massive star forming regions. In the following analysis, it can be noted that all of the six molecular clouds have distinct dust temperature gradient, ranging from $\sim$ 20 K to $\sim$ 30 K. This makes them good candidates to study chemical evolution of N$_2$H$^+$ in different evolutionary stages of massive star formation. From figure 1 to figure 6, we show the composed Spitzer images of our sources overlaid with the ATLASGAL 870 $\mu$m contours.
\subsection{MALT90}
We use archival molecular data from the MALT90 Survey. MALT90 is an international project with the aim to characterize physical and chemical properties of massive star formation in our Galaxy (e.g., Foster et al. 2011; Foster et al. 2013; Jackson et al. 2013). This project was carried out with the Mopra Spectrometer (MOPS) arrayed on the Mopra 22 m telescope, which is located near Coonabarabran in New South Wales, Australia. The full 8GHz bandwidth of MOPS was split into 16 zoom bands of 138 MHz, providing a velocity resolution of 0.11 km s$^{-1}$ at frequencies near 90 GHz. The beamsize of Mopra is 38$^\prime$$^\prime$ at 86GHz, with a beam efficiency between 0.49 at 86 GHz and 0.42 at 115 GHz (Ladd et al. 2005). The target of this survey are selected from the ATLASGAL clumps found by Contreras et al. (2013). The size of the data cube is 4.6$^\prime$ $\times$ 4.6$^\prime$, with a step of 9$^\prime$$^\prime$. We downloaded the data files from the MALT90 Home Page\footnote{http://atoa.atnf.csiro.au/MALT90}. For each source, we combined all the N$_2$H$^+$ data cube into a new data cube using the software package of CLASS (Continuum and Line Analysis Single-Disk Software). The analysis of other molecules (C$_2$H, HC$_3$N, H$^{13}$CO$^+$ and so on) will come in another paper. The combined images of N$_2$H$^+$ integrated emissions are also shown in figure 1-6.  An example of N$_2$H$^+$ (1-0) averaged spectrum is shown in figure 7.
\subsection{ATLASGAL and Hi-GAL}
ATLASGAL is the first systematic survey of the inner Galactic plane in the submillimetre (Siringo et al. 2009; Contreras et al. 2013). It provides high angular resolution ($\sim$ 19.2 arcsec) of cold dust emissions in the Galaxy. For a dust temperature of 20 K, ATLASGAL is sensitive to gas with H$_2$ column densities exceeding 10$^{22}$ cm$^{-2}$. The Hi-GAL data set is comprised of 5 continuum images of the Milky Way Galaxy using the PACS (70 and 160 $\mu$m) and SPIRE (250, 350 and 500 $\mu$m) instruments. The nominal angular resolutions ranges from 5.2$^\prime$$^\prime$ to 35.2$^\prime$$^\prime$ for 70 $\mu$m and 500 $\mu$m. The high-frequency components provide high angular resolution and is unaffected by large-scale background and foreground emissions, while the low-frequency used to maintain contaminations. Compared to ATLASGAL, Hi-GAL is more sensitive to dust emissions from low-density ISM. The two surveys provide us a unique unbiased catalogue of filament candidates in the Galaxy (e.g. Li et al. 2016; Andr\'e et al. 2010).

\section{Analysis}
\subsection{ Dust temperature and H$_2$ column density }
We made H$_2$ column density and dust temperature maps of each region by the spectral energy distribution (SED) method described by Wang et al. (2015). Given Hi-GAL is sensitive to low-density gas of about 10$^{21}$ cm$^{-2}$, background and/or foreground contaminations make a serious problem when analysing the Hi-GAL data. Following the steps described by Wang et al. (2015), we first remove the background and foreground emissions. After removing the background and foreground emissions, we re-gridded the pixels onto the same scale of 13$^\prime$$^\prime$, and convolved all the images to a spatial resolution of 45$^\prime$$^\prime$ which is the measured beamsize of Hi-GAL observations at 500 $\mu$m (Traficante et al. 2011). For each pixel, we use equation
\begin{equation}
I_\nu = B_\nu (1 - e^{-\tau_\nu })
\end{equation}
to model intensities at various wavelengths. The optical depth $\tau_\nu$ could be estimated through
\begin{equation}
\tau_\nu = \mu_{H_2} m_H \kappa_\nu N_{H_2} / R_{gd}
\end{equation}
We adopt a mean molecular weight per H$_2$ molecule of $\mu_{H_2}$ = 2.8 to include the contributions from Helium and other heavy elements. $m_H$ is the mass of a hydrogen atom. $N_{H_2}$ is the column density. R$_{gd}$ is the gas-to-dust mass ratio which is set to be 100. According to Ossenkopf $\&$ Henning (1994), dust opacity per unit dust mass ($\kappa_\nu$) could be expressed as
\begin{equation}
\kappa_\nu = 5.0 (\frac{\nu}{600 GHz})^\beta cm^{2} g^{-1}
\end{equation}
where the value of the dust emissivity index $\beta$ is fixed to 1.75 in our fitting. The two free parameters ($N_{H_2}$ and $T_d$) for each pixel could be fitted finally. The final resulting dust temperature and column density maps, which have a spatial resolution of 45$^\prime$$^\prime$ with a pixel size of 13$^\prime$$^\prime$, are shown in figure 8.

\subsubsection{Column density and abundance of N$_2$H$^+$}
To calculate the abundance of N$_2$H$^+$ in each pixel, we also smoothed the molecular data into a new beamsize of 45$^\prime$$^\prime$ with a new step of 13$^\prime$$^\prime$. Assuming LTE conditions and a beam filling factor of 1, the column density of N$_2$H$^+$ in every pixel can thus be calculated through:
\begin{equation}
N(N_2H^+) = \frac{8 \pi \nu^3}{c^3 R} \frac{Q_{rot}}{g_u A_{ul}}
\frac{exp(E_l/k T_{ex})}{1 - exp (-h \nu /k T_{ex})} \int\tau d\nu
\end{equation}
where $c$ is the velocity of light in the vacuum, $\nu$ is the frequency of the transitions, $g_u$ is the statistical weight of the upper level, $A_{ul}$ is the Einstein coefficient, $E_l$ is the energy of the lower level, $Q_{rot}$ is the partition function. For the excitation temperature of T$_{ex}$, we here assume that T$_{ex}$ is equal to the dust temperature derived above in each pixel. The value of R is 5/9, taking into account the satellite lines corrected by their relative opacities (Sanhueza et al. 2012). We use approximation
\begin{equation}
\int\tau d\nu = \frac{\tau}{1 - exp(-\tau)} \frac{\int T_{mb} dv}{J(T_{ex}) - J(T_{bg})}
\end{equation}
to take $\tau_{N_2H^+}$ into account. N$_2$H$^+$ (1-0) has 7 hyperfine components (e.g., Pagani et al. 2009; Keto $\&$ Rybicki 2010). As an example is shown in figure 7, the 7 hyperfine structures of N$_2$H$^+$ (1-0) blended into 3 groups because of turbulent line widths. We estimate the optical depth of N$_2$H$^+$ (1-0) by the method described by Purcell et al.(2009). The integrated intensities of Group 1/Group 2 should be in the ratio of 1:5, assuming that the line widths of the individual hyperfine components are all equal. The optical depth of N$_2$H$^+$ ($\tau_{N_2H^+}$) can then be derived using the following equation:
\begin{equation}
\frac{\int T_{MB,Group1} dv}{\int T_{MB,Group2} dv} = \frac{1 -
exp(-0.2\tau_2)}{1 - exp(-\tau_2)}
\end{equation}
By solving Equation (4) and (6) in each pixel where the Group 2 emission of N$_2$H$^+$ is greater than 6 $\sigma$, we got the column density maps of N(N$_2$H$^+$). For the uncertainties of column density, we here only consider the errors from optical depth and integrated intensities. The mean uncertainty is about 20$\%$. The abundance value of N$_2$H$^+$ ($\chi$(N$_2$H$^+$)) for each pixel can be calculated through $\chi$(N$_2$H$^+$) = N(N$_2$H$^+$)/N(H$_2$). The N$_2$H$^+$ abundance maps are shown in figure 9.

\section{Results and Discussions}
\subsection{G351.776-0.527}
G351.776-0.527 is a filamentary cloud identified by Li et al. (2016). It is also known as infrared dark cloud (IRDC) G351.77-0.51 (Simon et al. 2006). In the top panel of figure 1, dark absorption features against the Galactic mid-infrared background radiation field are obvious. According to Leurini et al. (2011), the kinematic distance of this cloud is about 1 kpc. IRAS 17233-3606 lies in the center of the filament, where the dust temperature is more than 30 K. Previous studies indicate active massive star formation in this IRAS source (e.g. Caswell et al. 1980; Menten 1991; Leurini et al. 2009). Using high resolution observations of VLA, Klaassen et al. (2015) found a large scale outflow from IRAS 17233-3606. Some dense clumps have also found on the northeast and southwest part of G351.776-0.527. The dust temperature there is relatively low, indicating earlier evolutionary stages. From the top left panel of figure 9, we can see that the abundance of N$_2$H$^+$ is highest in the center and southeast side of IRAS 17233-3606. The top left panel of figure 10 indicates the abundance of N$_2$H$^+$ increase as the increment of dust temperature in G351.776-0.527. This result is inconsistent with chemical models of low-mass star formation.

\subsection{G340.301-0.387}
G340.301-0.387 is also a filamentary cloud identified by Li et al. (2016). From figure 2, we can see that most of the dense gas is in the northwest part of this cloud, where the gas is also more evolved. The clumps in the southeast show no distinct Spitzer 8 $\mu$m emissions, indicating the gas here is less evolved. The dust temperature map shows a distinct temperature gradient, decreasing from $\sim$ 27 K in the northwest to $\sim$ 20 K in the southeast. The N$_2$H$^+$ abundance map shows a similar trend, also decreasing from northwest to southeast. The T$_d$ - $\chi$(N$_2$H$^+$) relation map in figure 10 suggests a positive correlation. This trend is also inconsistent with chemical models of low-mass star formation, which suggests the abundance of N$_2$H$^+$ should decrease as star formation evolves.

\subsection{CN148 and S36}
CN148 and S36 are two infrared bubbles found by Churchwell et al. (2006; 2007). The arc-shaped distributions of N$_2$H$^+$ (1-0) emission, the 870 $\mu$m dust emission, and the polycyclic aromatic hydrocarbon (PAH) features trace photodissociation regions (PDRs) around the two bubbles. A multi-wavelength study of CN148 carried out by Dewangan et al. (2015) suggests triggered star formation when the bubble expands into the surrounding interstellar medium (ISM). A distinct dust temperature gradient can be noted around the two bubbles. The dust is quite hot (more than 30 K) on the PDRs, and decreases to $\sim$ 20 K outside of PDRs. According to Guzm\'{a}n et al. (2015), the clumps found by Contreras et al. (2013) in the two regions range from the ``Quiescent" to PDR stages. Our study shows the abundance of N$_2$H$^+$ is relatively low on the PDRs. This is consistent with chemical models, as N$_2$H$^+$ is regarded to be destroyed by free electrons (e.g. Dislaire et al. 2012; Vigren et al. 2012). In a previous paper (Yu $\&$ Xu 2016), we also found the abundance of N$_2$H$^+$ seems to decrease as a function of Lyman continuum fluxes (N$_L$) in compact HII regions, indicating that this molecule could be destroyed by UV photons when H II regions have formed. The T$_d$ - $\chi$(N$_2$H$^+$) relation maps in figure 10 suggest the abundance of N$_2$H$^+$ increases when the dust temperature increases from $\sim$ 18 K to $\sim$ 27 K, and drops when dust temperature is more than 27 K. Again, the evolution trend of N$_2$H$^+$ is inconsistent with chemical models when dust temperature is below 27 K.

\subsection{G326.432+0.916 and G326.641+0.612}
These two sources are two dense clouds we selected from Rathborne et al. (2016). Even though they are not listed as bubbles by Churchwell et al. (2006; 2007), we can see distinct radio emissions from the Sydney University Molonglo Sky Survey (SUMSS) (843 MHz; Mauch et al. 2003), indicating they are also two classical HII regions. The situation of these two sources are quite similar to CN148 and S36. The dust temperature in the two regions also decreases from PDR to the gas outside. The T$_d$ - $\chi$(N$_2$H$^+$) relation maps of the two sources also show a turning point near 27 K.

\subsection{Discussions}
In the chemical models of low-mass star formation (e.g. Bergin $\&$ Langer 1997; Lee et al. 2004), a relative enhancement of N$_2$H$^+$ abundance is expected in the cold prestellar phase, as CO is thought to be depleted in starless cores. As the central star evolves, the gas gets warm and CO should evaporate from the dust grains if the dust temperature exceeds about 20 K (Tobin et al. 2013). Thus we could expect the N$_2$H$^+$ abundance decrease as a function of dust temperature. In order to investigate the chemical evolution of N$_2$H$^+$ as clumps evolve, here we present a multiwavelength study toward six molecular clouds containing several clumps in different evolutionary stages of star formation. The initial conditions could be supposed to be the same in the same molecular cloud. Our study indicates when dust temperature is below 27 K, the abundance of N$_2$H$^+$ increases with dust temperature. Previous studies (e.g. Hoq et al. 2013, Sanhueza et al. 2012, Miettinen 2014) also show this similar trend. The result of our study is consistent with those previous studies, although we used a different approach. As Hoq et al. (2013) suggest, chemical processes in massive star-forming regions may really differ from low-mass star formation. The mass infall rate, UV flux, and density in massive star formation regions are indeed different from their low-mass counterparts. The large beam of Mopra may also be the reason for this inconsistency. Chemical models of low-mass star forming regions (e.g. Bergin $\&$ Langer 1997; Lee et al. 2004) focused on single low-mass cores, which cannot be compared with a clump. Higher angular resolution observations and chemical models should be carried out to study the chemical evolution of N$_2$H$^+$ in the early stages of massive star formation. In the PDRs where dust temperature is more than 27 K, our study indicates that the abundance of N$_2$H$^+$ begins to decrease (CN148, S36 and G326.432+0.916) or reaches a plateau (G351.776-0.527 and G326.641+0.612). This is consistent with chemical models, as N$_2$H$^+$ is prone to be destroyed by free electrons (e.g. Vigren et al. 2012; Yu \& Xu 2016). Figure 14 in Sanhueza et al. (2012) and Figure 5 in Hoq et al. (2013) indicate the increase of N$_2$H$^+$ abundance was up to the protostellar phase, and then for the ``red clumps" or ``PDR clumps" there was no significant increase or decrease. This phenomena is very similar to our sources of G351.776-0.527 and G326.641+0.612, where the N$_2$H$^+$ abundance reaches a plateau and becomes more or less constant when T$_d$ $>$ 27 K. Hoq et al. (2013) do not see the decrease from their ``Protostellar" to ``HII/PDR" clumps, because most sources in their ``HII/PDR" catalogue are compact HII regions, which means the size of ionized gas is no more than 0.1 pc. Given a typical distance of 3 kpc for a massive star forming region, the Mopra telescope not only probes the ionized gas but also the surrounding cold diffuse material. Large scale infalls have also been found in many compact HII regions (e.g. Keto $\& Wood$ 2006; Yu et al. 2015). In our six sources, all the PDRs locate around classical HII regions, where there is no large scale infall and the ionized gas could be resolved by the Mopra telescope. This may be the reason that we see the decrease of $\chi$(N$_2$H$^+$) when T$_d$ is above 27 K in CN148, S36 and G326.432+0.916.

\section{Summary}
We present a multi-wavelength study toward six massive star-forming regions, to investigate the chemical evolution of N$_2$H$^+$ as clumps evolve. Using archival data of Hi-GAL, we made H$_2$ column density and dust temperature maps of these regions through the SED method. We found all of the six sources show distinct dust temperature gradient, ranging from $\sim$ 20 K to $\sim$ 30 K. Previous infrared studies and the dust temperature images indicate physical and chemical properties are quite different in different parts of these sources. This makes them good candidates for us to study the chemical evolution of N$_2$H$^+$ in different evolutionary stages of massive star formation. Using molecular line data of MALT90, we made the abundance maps of N$_2$H$^+$. We found that when the dust temperature is above 27 K, the abundance of N$_2$H$^+$ begins to decrease or reaches a plateau. We regard this is because that in the PDRs around classical HII regions, N$_2$H$^+$ is destroyed by electrons heavily. We also found that when the dust temperature is below 27 K, the abundance of N$_2$H$^+$ increases with dust temperature. This is inconsistent with chemical models of low-mass star formation. In order to check out whether this inconsistency is caused by a different chemistry in high-mass star-forming clumps, higher angular resolution observations are necessary.

\section{ACKNOWLEDGEMENTS}
We are very grateful to the anonymous referee for his/her helpful comments and suggestions. This paper has made use of information from the ATLASGAL Database Server\footnote{ http://atlasgal.mpifr-bonn.mpg.de/cgi-bin/ATLASGAL$_-$DATABASE.cgi}. The Red MSX Source survey was constructed with support from the Science and Technology Facilities Council of the UK. The ATLASGAL project is a collaboration between the Max-Planck-Gesellschaft, the European Southern Observatory (ESO) and the Universidad de Chile. This research made use of data products from the Millimetre Astronomy Legacy Team 90 GHz (MALT90) survey. The Mopra telescope is part of the Australia Telescope and is funded by the Commonwealth of Australia for operation as National Facility managed by CSIRO. This
paper is supported by National Natural Science Foundation of China under grants of 11503037.

\begin{deluxetable}{cccccccc}
\tablewidth{0pt} \tablecaption{ Basic Information of our sources.}
\renewcommand{\arraystretch}{0.8}
\tablehead{
Source & Dist.$^{a}$ & Associated   &  $l$  & $b$  & R$_{eff}$$^b$ &Mass$^c$  & Type$^b$ \\
Name & (kpc) & Clumps & (deg) & (deg) & (pc) &(M$_{\odot}$) & \\
}\startdata

G351.776-0.527 & 1.0   \\
&&AGAL351.774-00.537 & 351.774 & -0.537 & 0.37 & 377.6 & Protostellar \\
&&AGAL351.784-00.514 & 351.784 & -0.514 & 0.34 & 461.4 & Protostellar \\
&&AGAL351.744-00.577 & 351.744 & -0.577 & 0.44 & 431.0 & HII region \\
&&AGAL351.783-00.604 & 351.783 & -0.604 & 0.15 & 40.8  & Quiescent \\
&&AGAL351.804-00.449 & 351.804 & -0.449 & 0.38 & 392.4 & Protostellar \\
    \hline\\
 G340.301-0.387 & 3.9\\
&&AGAL340.248-00.374 & 340.248 & -0.374 & 1.77 & 6121.9 & HII region \\
&&AGAL340.232-00.397 & 340.232 & -0.397 & 0.55 & 220.7 & Quiescent \\
&&AGAL340.304-00.376 & 340.304 & -0.376 & 0.89 & 1077.0 & Quiescent \\
&&AGAL340.301-00.402 & 340.301 & -0.402 & 0.81 & 1130.1 & Quiescent \\
    \hline\\
 CN148 & 2.2\\
&&AGAL010.299-00.147 & 10.299 & -0.147 & 0.70 & 923.3 & HII region \\
&&AGAL010.286-00.164 & 10.286 & -0.164 & 0.49 & 285.4 & Quiescent \\
&&AGAL010.323-00.161 & 10.323 & -0.161 & 0.74 & 837.3 & HII region \\
&&AGAL010.284-00.114 & 10.284 & -0.114 & 0.37 & 615.7 & Quiescent \\
&&AGAL010.342-00.142 & 10.342 & -0.142 & 0.44 & 393.0 & Protostellar \\
&&AGAL010.344-00.172 & 10.344 & -0.172 & 0.38 & 184.7 & PDR \\
&&AGAL010.356-00.149 & 10.356 & -0.149 & 0.53 & 480.8 & HII region \\
&&AGAL010.404-00.201 & 10.404 & -0.201 & 0.44 & 397.2 & HII region \\
    \hline\\
 S36 & 3.2 \\
&&AGAL337.922-00.456 & 337.922 & -0.456 & 1.27 & 3001.0 & PDR \\
&&AGAL337.916-00.477 & 337.916 & -0.477 & 0.93 & 1927.5 & Protostellar \\
&&AGAL337.934-00.507 & 337.934 & -0.507 & 0.93 & 1153.7 & PDR \\
&&AGAL337.939-00.532 & 337.939 & -0.532 & 0.53 & 400.3 & HII region\\
&&AGAL337.974-00.519 & 337.974 & -0.519 & 0.60 & 195.4 & PDR\\
    \hline\\
 G326.432+0.916 & 2.9 \\
&&AGAL326.446+00.907 & 326.446 & 0.907 & 0.92 & 510.4 & HII region \\
&&AGAL326.407+00.927 & 326.407 & 0.927 & 0.84 & 1059.6 & Protostellar \\
&&AGAL326.472+00.889 & 326.472 & 0.889 & 1.19 & 2790.3 & Uncertain \\
&&AGAL326.491+00.882 & 326.491 & 0.882 & 0.53 & 737.1 & Uncertain \\

    \hline\\
 G326.641+0.612 & 2.8 \\
&&AGAL326.641+00.612 & 326.641 & 0.612 & 0.94 & 5557.1 & PDR \\
&&AGAL326.627+00.611 & 326.627 & 0.611 & 1.05 & 4280.1 & PDR \\
&&AGAL326.657+00.594 & 326.657 & 0.594 & 1.01 & 1922.4 & UCHII region\\
&&AGAL326.681+00.564 & 326.681 & 0.564 & 0.64 & 828.1 & Uncertain\\
&&AGAL326.671+00.554 & 326.671 & 0.554 & 0.54 & 601.1 & UCHII region\\
\enddata
\tablenotetext{a} {The distance value of G351.776-0.527 comes from Leurini et al. (2011). The distance value of CN148 comes from Dewangan et al. (2015). The distances of other sources come from Whitaker et al. (2017). }
\tablenotetext{b}{Guzm\'{a}n et al. (2015).}
\tablenotetext{c}{Contreras et al. (2017).}

\end{deluxetable}

\clearpage

%fig1
\begin{figure}
\center
\psfig{file=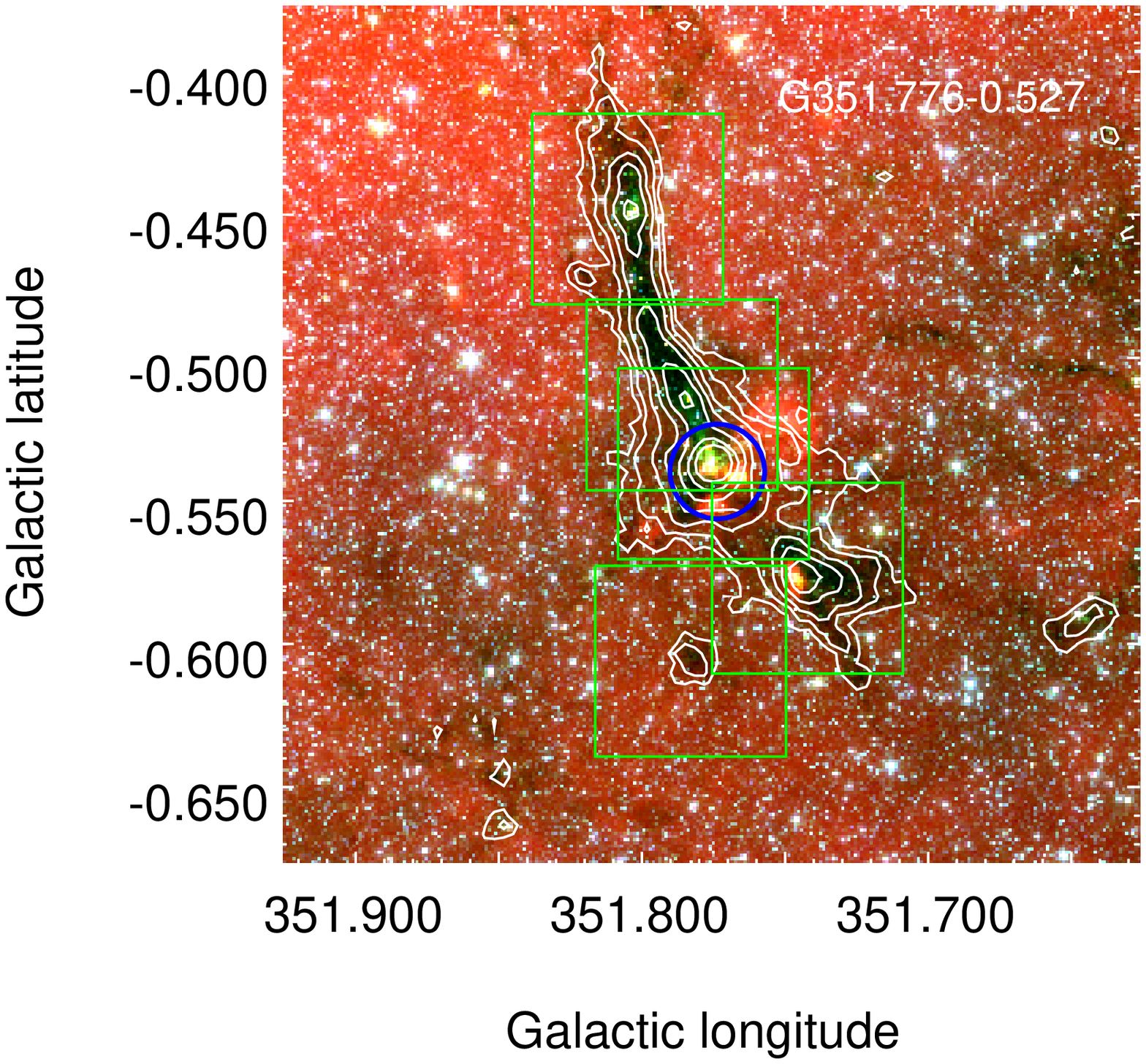,width=4in,height=3.3in}
\psfig{file=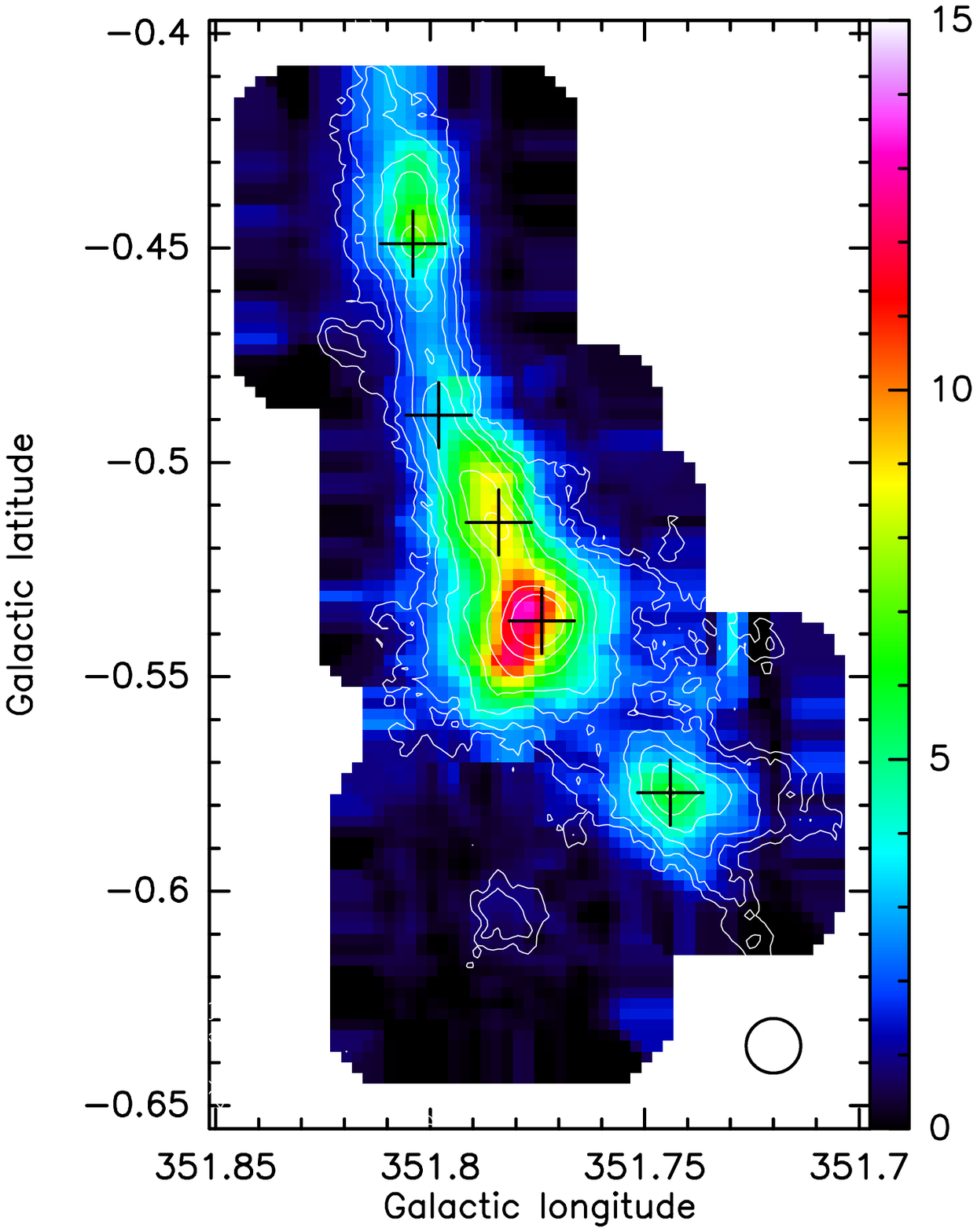,width=2.7in,height=3.5in} \caption{Top: Three colour mid-infrared image of G351.776-0.527 created using the Spitzer IRAC band filters (8.0 $\mu$m in red, 4.5 $\mu$m in green and 3.6 $\mu$m in blue). The green boxes indicate the five observed regions by MALT90. The blue circle marks IRAS 17233-3606. Bottom: The new combined image of N$_2$H$^+$ from MALT90 data set. The emission has been integrated from -6 to 0 km s$^{-1}$. The ATLASGAL 870 $\mu$m emissions (in white) are superimposed with levels 0.15, 0.30, 0.60, 1.20, 2.40, 4.80 and 9.60 Jy/beam in each panel. The pluses mark the dense clumps listed in table 1. The black circle shown in the lower right corner of this image indicates the beam size of Mopra. The unit of the color bar on the right is in K km/s. }
\end{figure}

%fig2
\begin{figure}
\center
\psfig{file=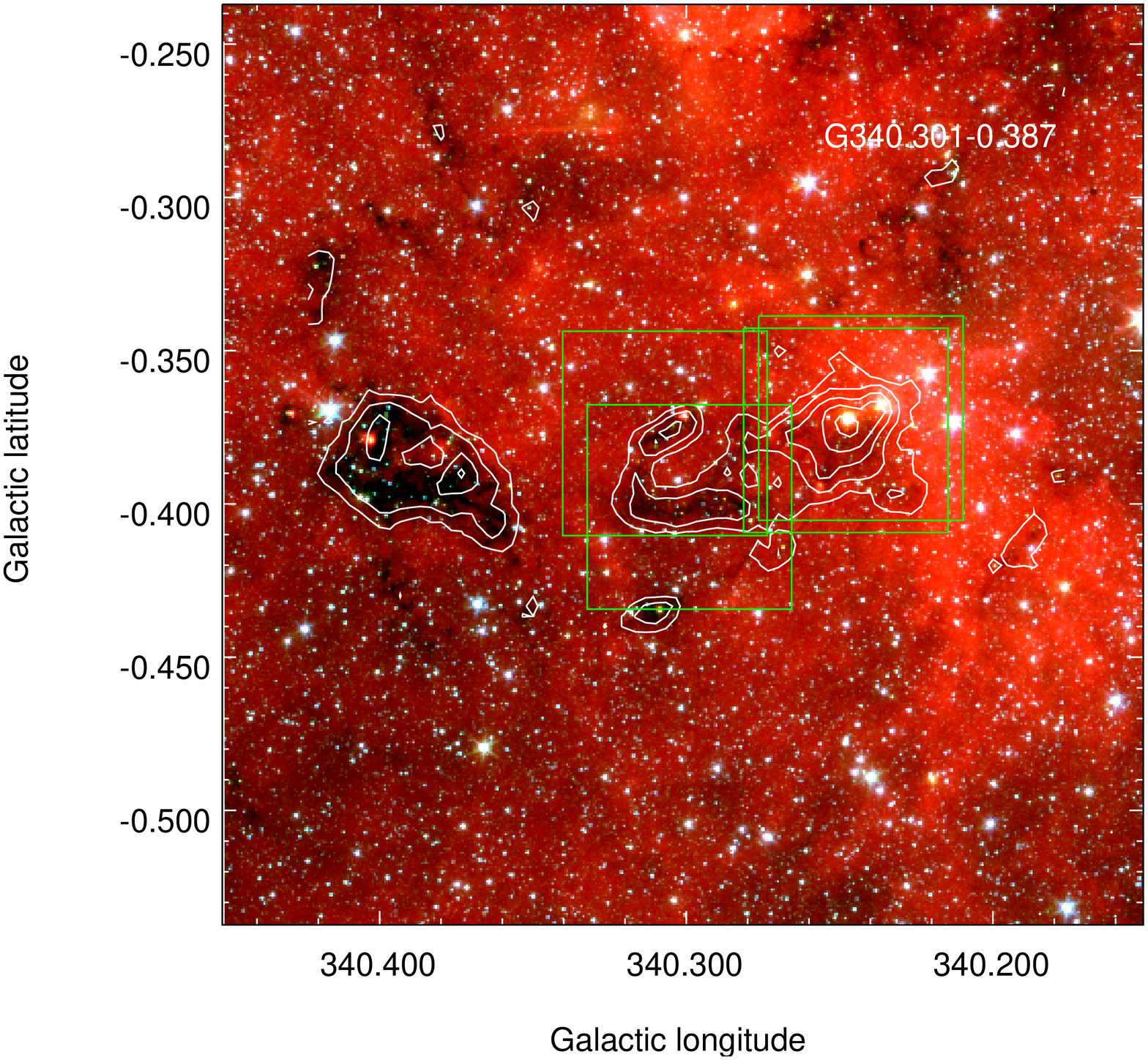,width=4in,height=3.0in}
\psfig{file=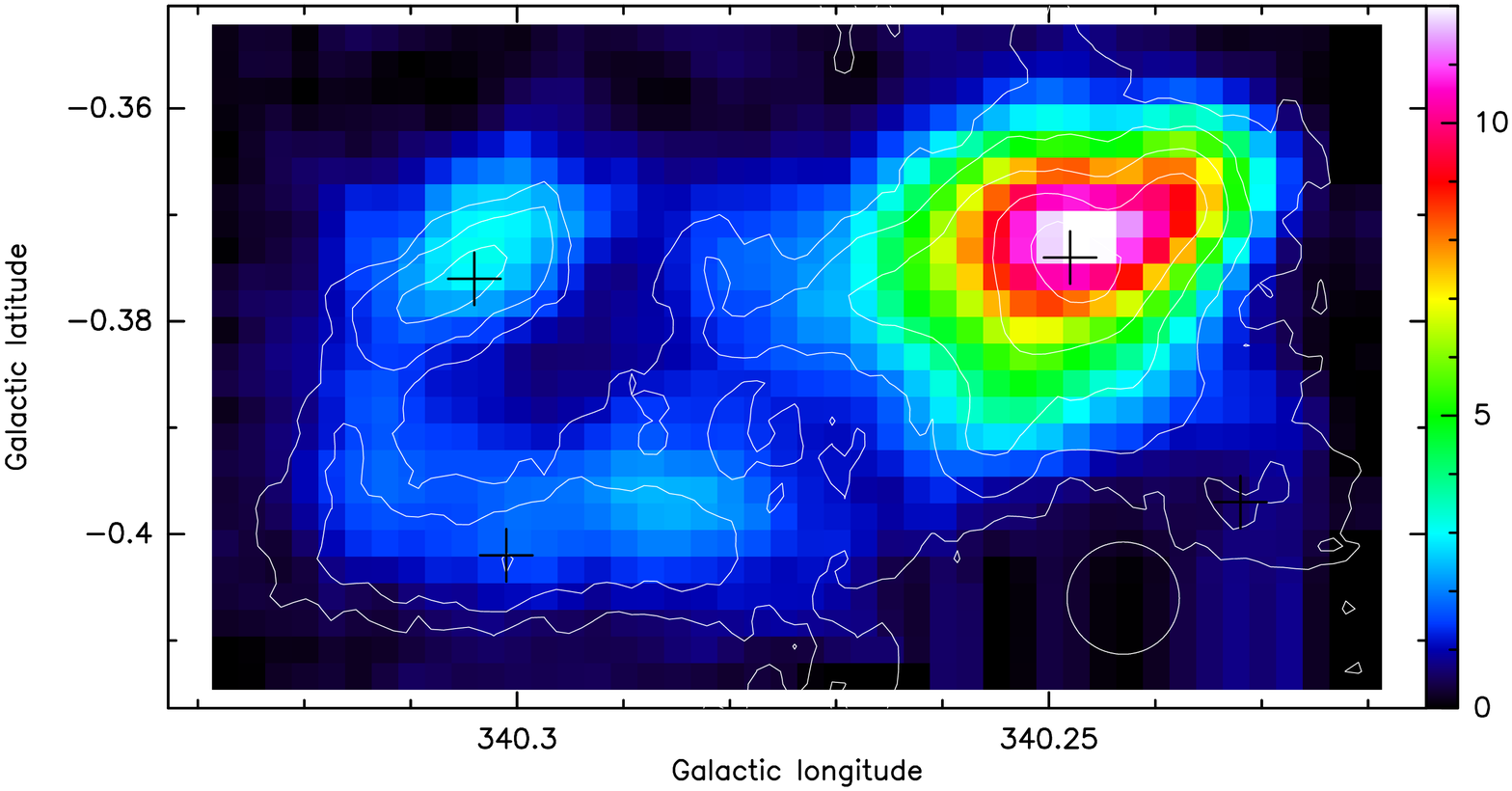,width=2.9in,height=1.5in} \caption{Top: Three colour mid-infrared image of \textbf{G340.301-0.387} created using the Spitzer IRAC band filters (8.0 $\mu$m in red, 4.5 $\mu$m in green and 3.6 $\mu$m in blue). The green boxes indicate the four observed regions by MALT90. Bottom: The new combined image of N$_2$H$^+$ from MALT90 data set. The emission has been integrated from -55 to -48 km s$^{-1}$. The ATLASGAL 870 $\mu$m emissions (in white) are superimposed with levels 0.24, 0.48, 0.96, 1.92 and 3.84 Jy/beam in each panel. The pluses mark the dense clumps listed in table 1. The white circle shown in the lower right corner of this image indicates the beam size of Mopra. The unit of the color bar on the right is in K km/s.}
\end{figure}

%fig3
\begin{figure}
\center
\psfig{file=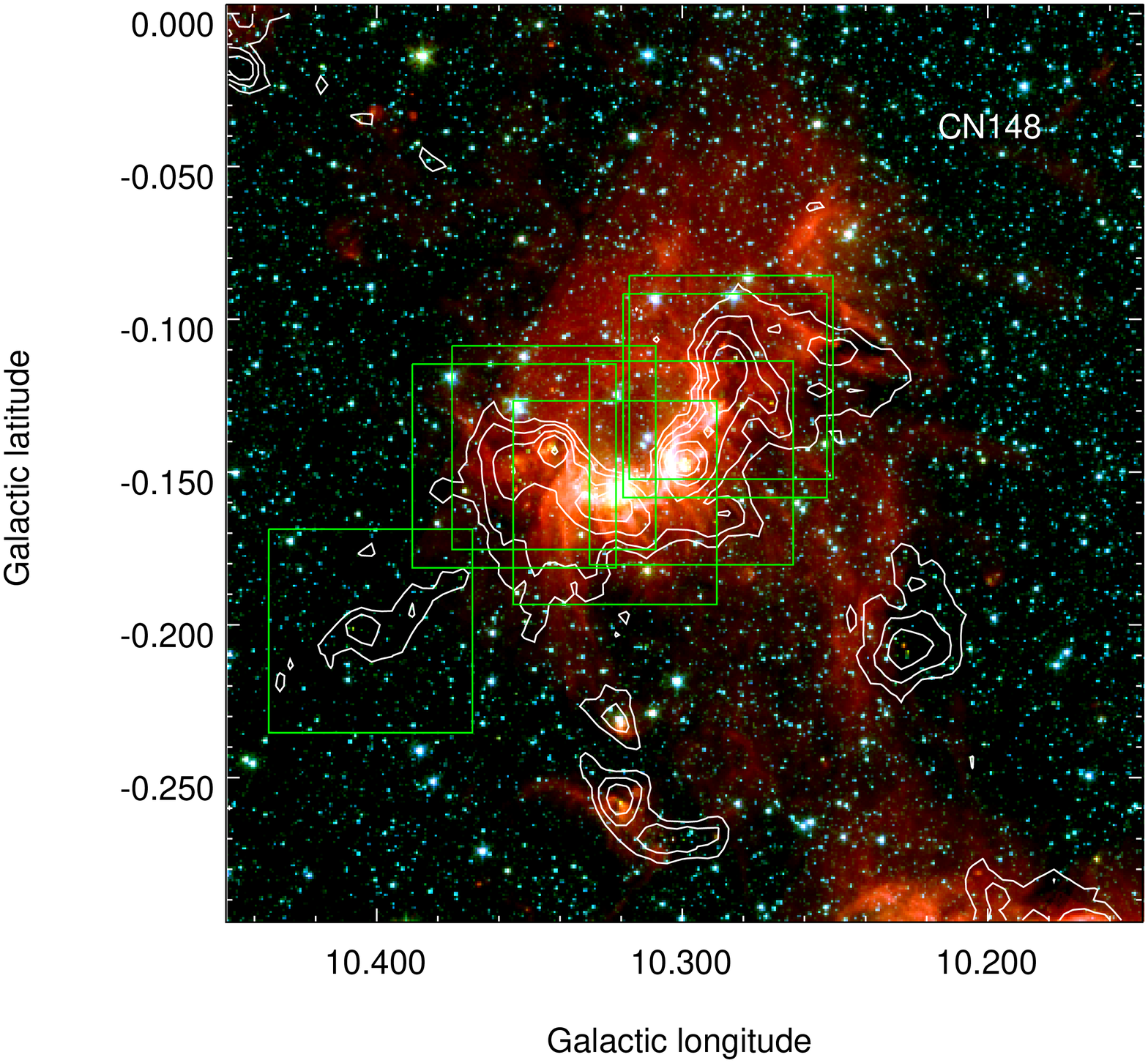,width=4in,height=3.0in}
\psfig{file=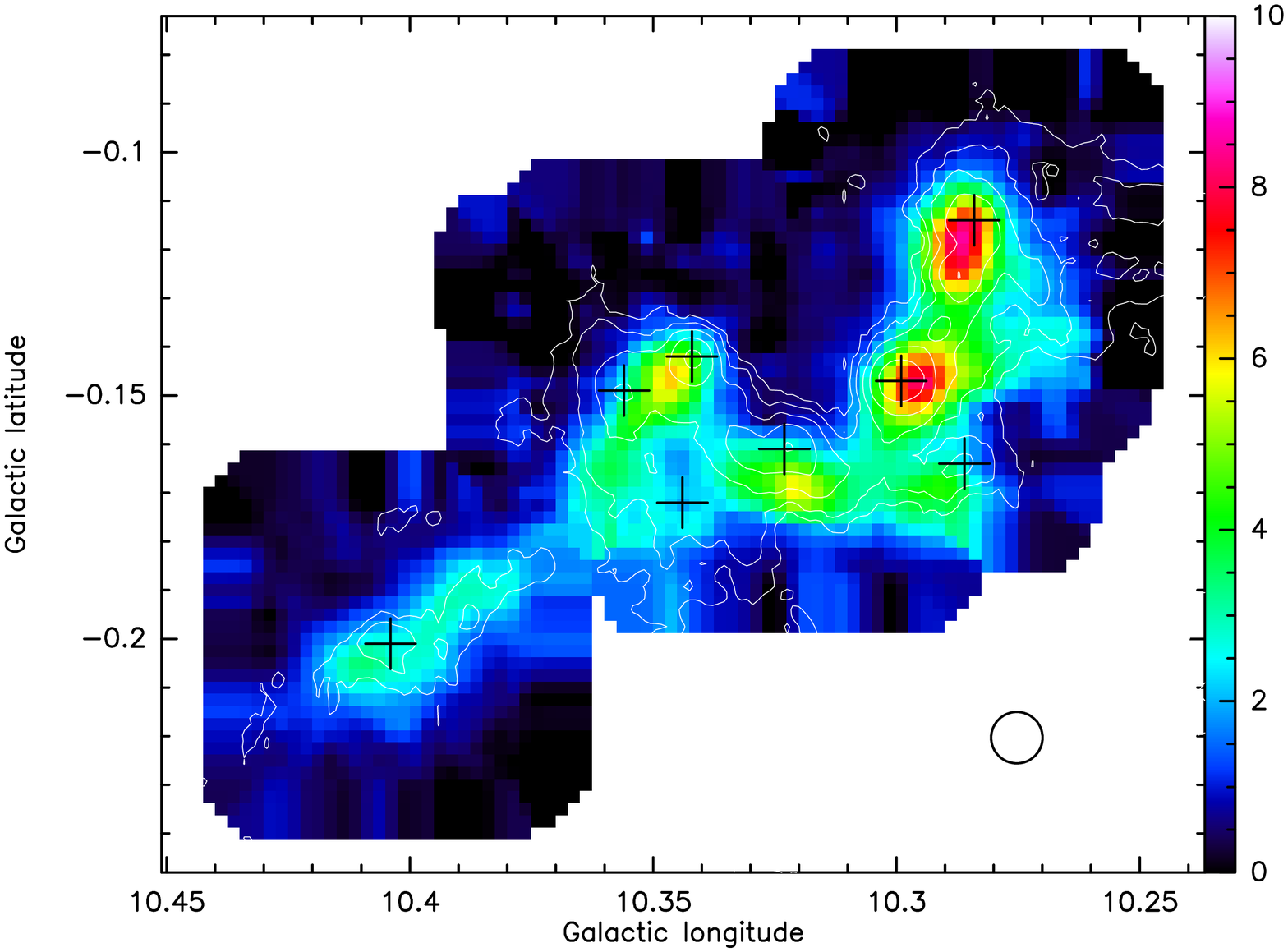,width=3.4in,height=2.5in} \caption{Top: Three colour mid-infrared image of CN148 created using the Spitzer IRAC band filters (8.0 $\mu$m in red, 4.5 $\mu$m in green and 3.6 $\mu$m in blue). The green boxes indicate the seven observed regions by MALT90. Bottom: The new combined image of N$_2$H$^+$ from MALT90 data set. The emission has been integrated from 10 to 15 km s$^{-1}$. The ATLASGAL 870 $\mu$m emissions (in white) are superimposed with levels 0.27, 0.54, 1.08, 2.16 and 4.32 Jy/beam in each panel. The pluses mark the dense clumps listed in table 1. The black circle shown in the lower right corner of this image indicates the beam size of Mopra. The unit of the color bar on the right is in K km/s.}
\end{figure}

%fig4
\begin{figure}
\center
\psfig{file=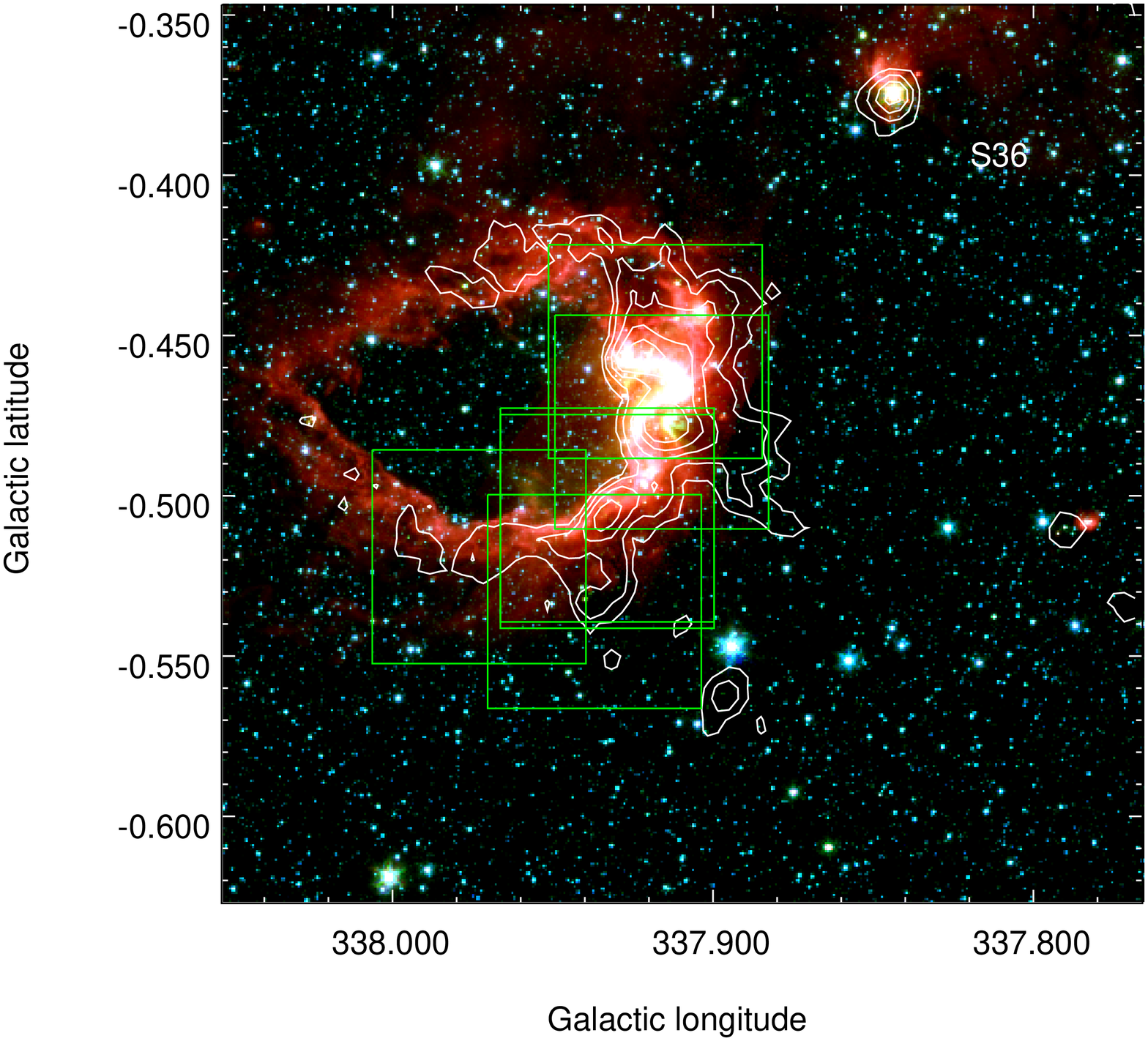,width=4in,height=3.0in}
\psfig{file=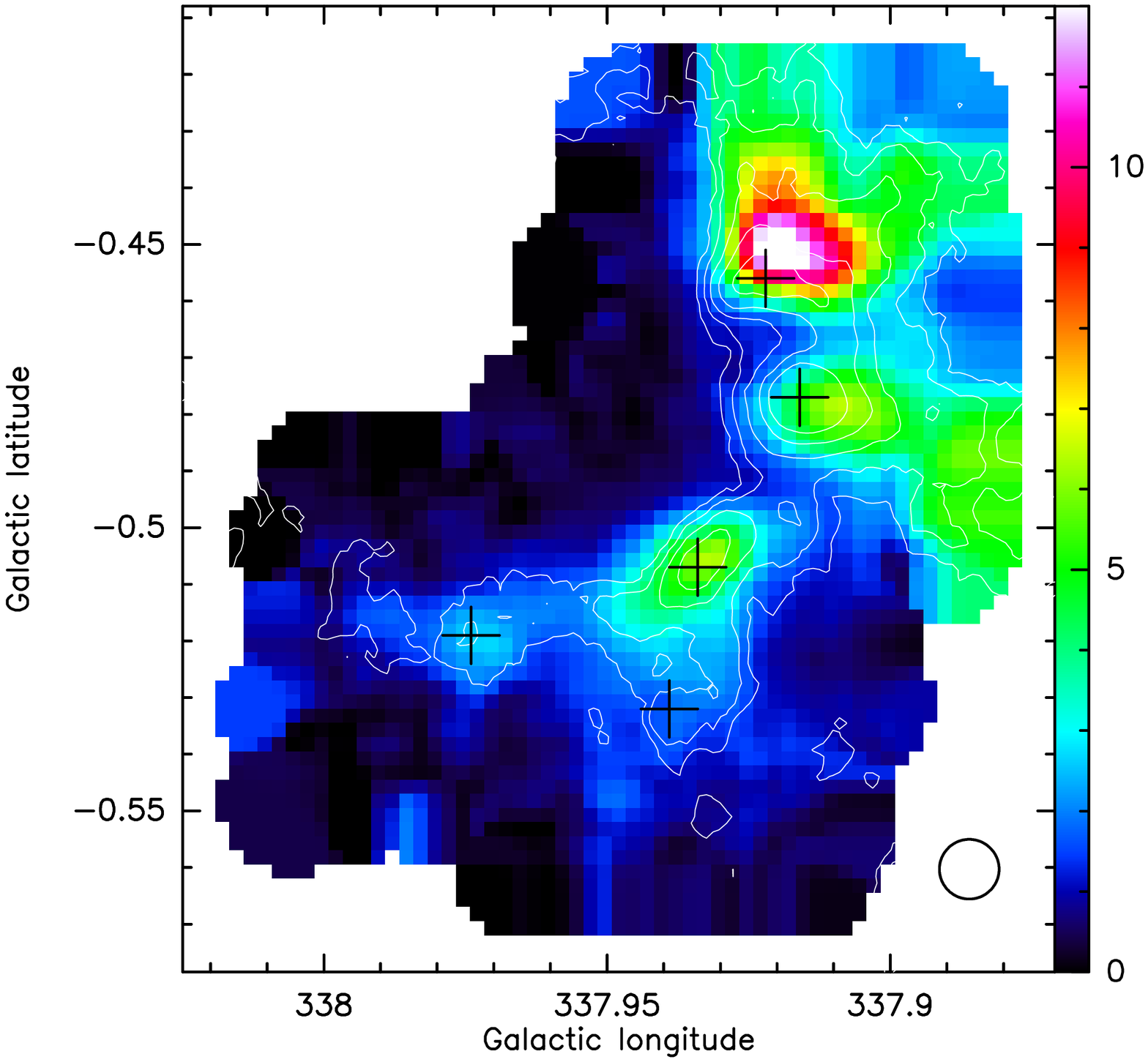,width=2.7in,height=2.5in} \caption{Top: Three colour mid-infrared image of S36 created using the Spitzer IRAC band filters (8.0 $\mu$m in red, 4.5 $\mu$m in green and 3.6 $\mu$m in blue). The green boxes indicate the six observed regions by MALT90. Bottom: The new combined image of N$_2$H$^+$ from MALT90 data set. The emission has been integrated from -42 to -36 km s$^{-1}$. The ATLASGAL 870 $\mu$m emissions (in white) are superimposed with levels 0.27, 0.54, 1.08, 2.16 and 4.32 Jy/beam in each panel. The pluses mark the dense clumps listed in table 1. The black circle shown in the lower right corner of this image indicates the beam size of Mopra. The unit of the color bar on the right is in K km/s.}
\end{figure}

%fig5
\begin{figure}
\center
\psfig{file=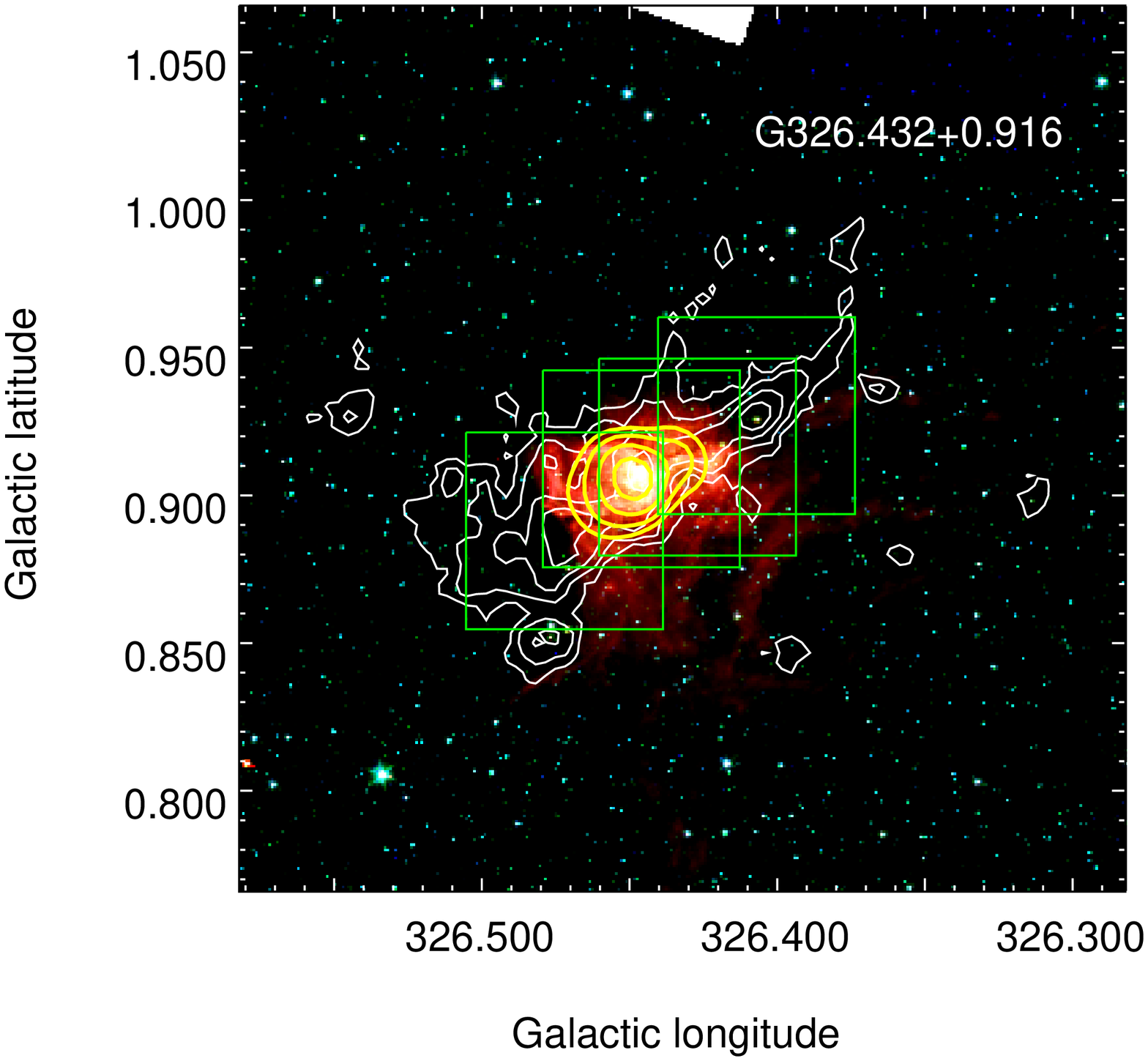,width=4in,height=3.0in}
\psfig{file=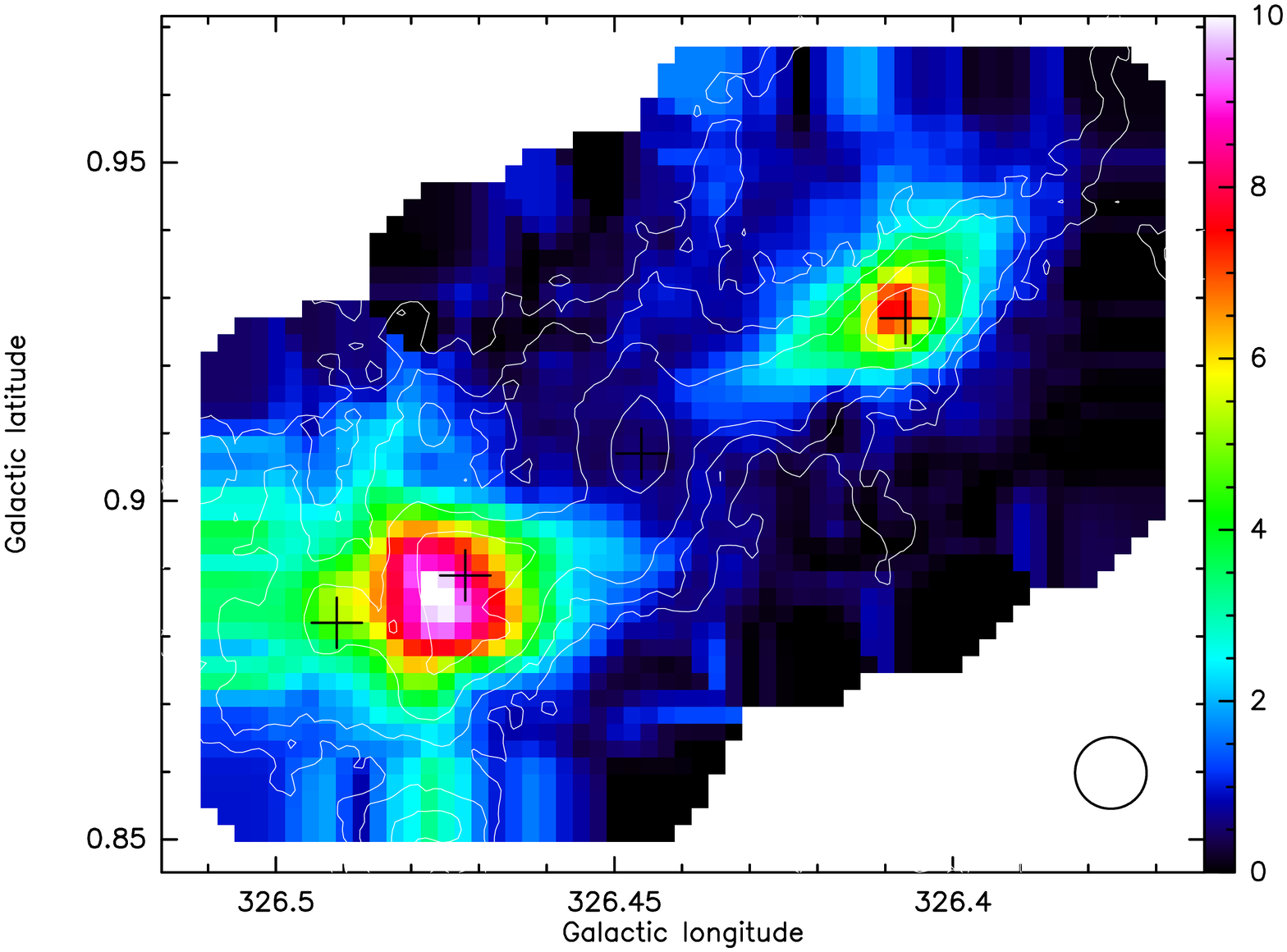,width=3in,height=2.2in} \caption{Top: Three colour mid-infrared image of G326.432+0.916 created using the Spitzer IRAC band filters (8.0 $\mu$m in red, 4.5 $\mu$m in green and 3.6 $\mu$m in blue). The green boxes indicate the four observed regions by MALT90. The 843 MHz SUMSS radio continuum emissions (yellow) are superimposed with levels 0.2, 0.4,0.8, 1.6 and 3.2 Jy/beam. Bottom: The new combined image of N$_2$H$^+$ from MALT90 data set. The emission has been integrated from -44 to -37 km s$^{-1}$. The ATLASGAL 870 $\mu$m emissions (in white) are superimposed with levels 0.21, 0.42, 0.84 and 1.68 Jy/beam in each panel. The pluses mark the dense clumps listed in table 1. The black circle shown in the lower right corner of this image indicates the beam size of Mopra. The unit of the color bar on the right is in K km/s.}
\end{figure}

%fig6
\begin{figure}
\center
\psfig{file=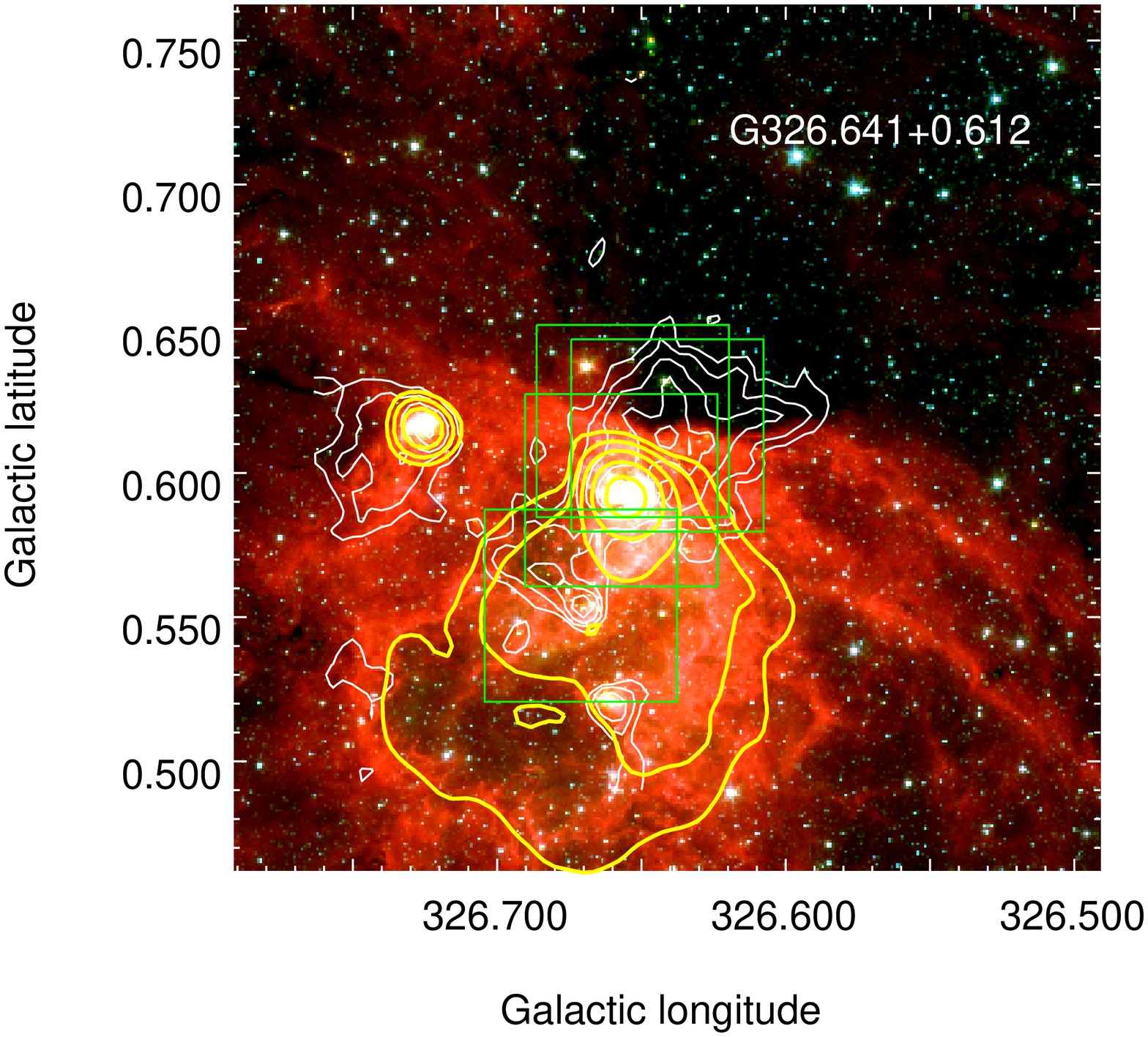,width=4in,height=3.5in}
\psfig{file=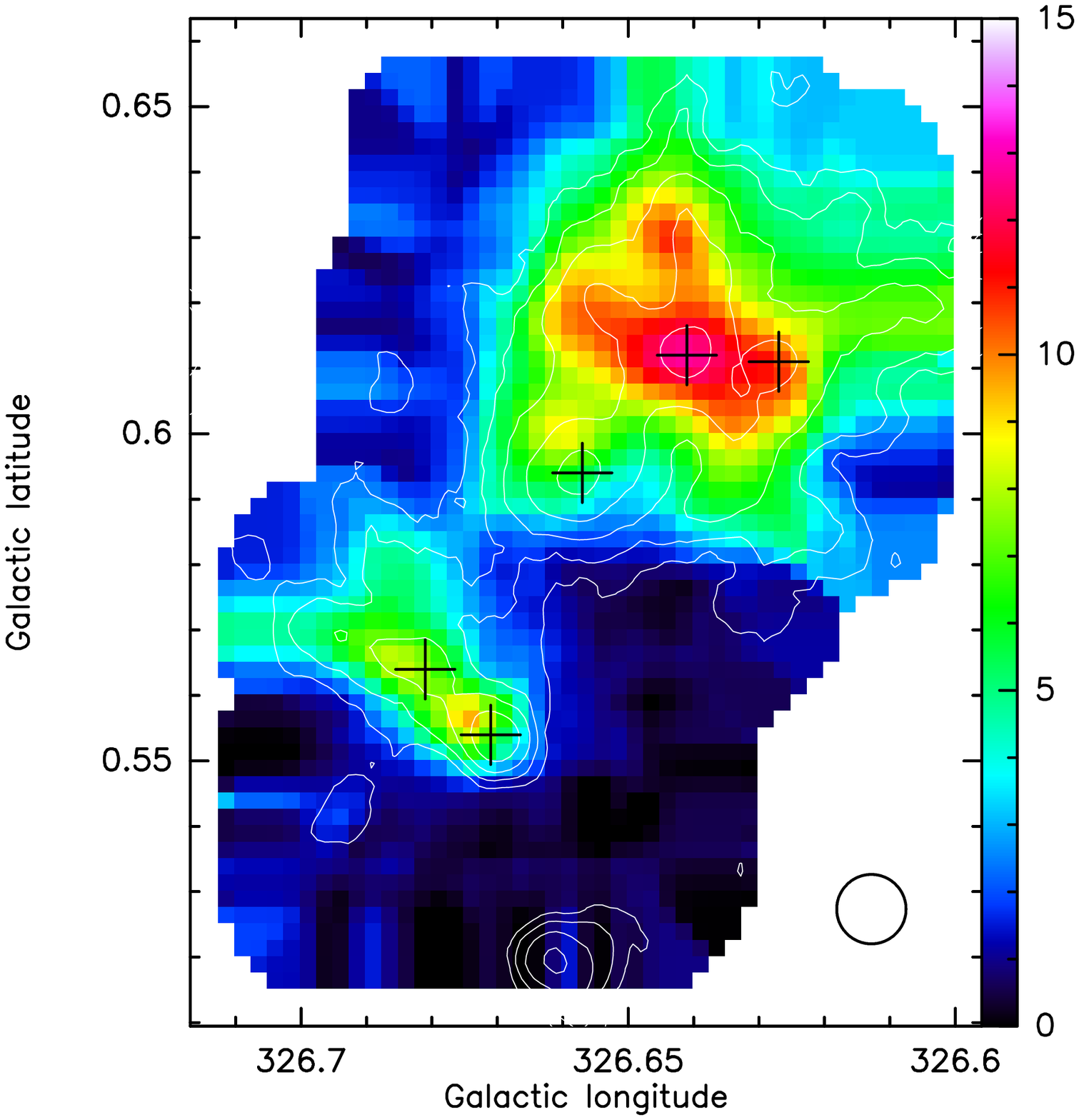,width=2.9in,height=3in} \caption{Top: Three colour mid-infrared image of G326.641+0.612 created using the Spitzer IRAC band filters (8.0 $\mu$m in red, 4.5 $\mu$m in green and 3.6 $\mu$m in blue). The green boxes indicate the four observed regions by MALT90. The 843 MHz SUMSS radio continuum emissions (yellow) are superimposed with levels 0.2, 0.4,0.8, 1.6 and 3.2 Jy/beam. Bottom: The new combined image of N$_2$H$^+$ from MALT90 data set. The emission has been integrated from -44 to -37 km s$^{-1}$. The ATLASGAL 870 $\mu$m emissions (in white) are superimposed with levels 0.42, 0.84, 1.68, 3.36 and 6.72 Jy/beam in each panel. The pluses mark the dense clumps listed in table 1. The black circle shown in the lower right corner of this image indicates the beam size of Mopra. The unit of the color bar on the right is in K km/s.}
\end{figure}

%fig7
\begin{figure}
\centering
\includegraphics[width=9cm, height=5.5cm]{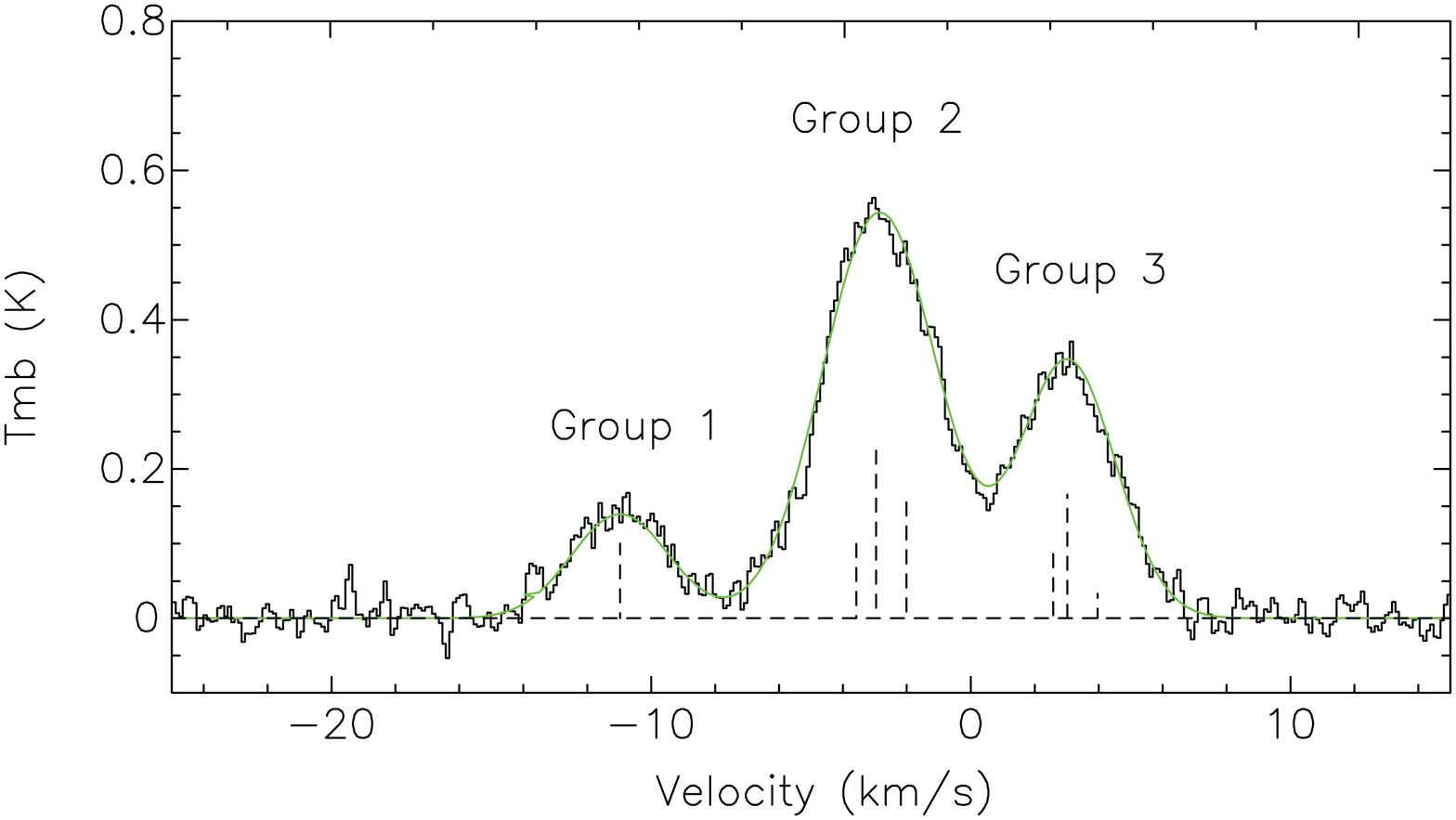}
\caption{Averaged spectra of N$_2$H$^+$ (1-0) over the cloud of G351.776-00.527. The vertical dashed lines indicate the seven hyperfine structures. }
\end{figure}

%fig8 to 13
\begin{figure}
\center
\psfig{file=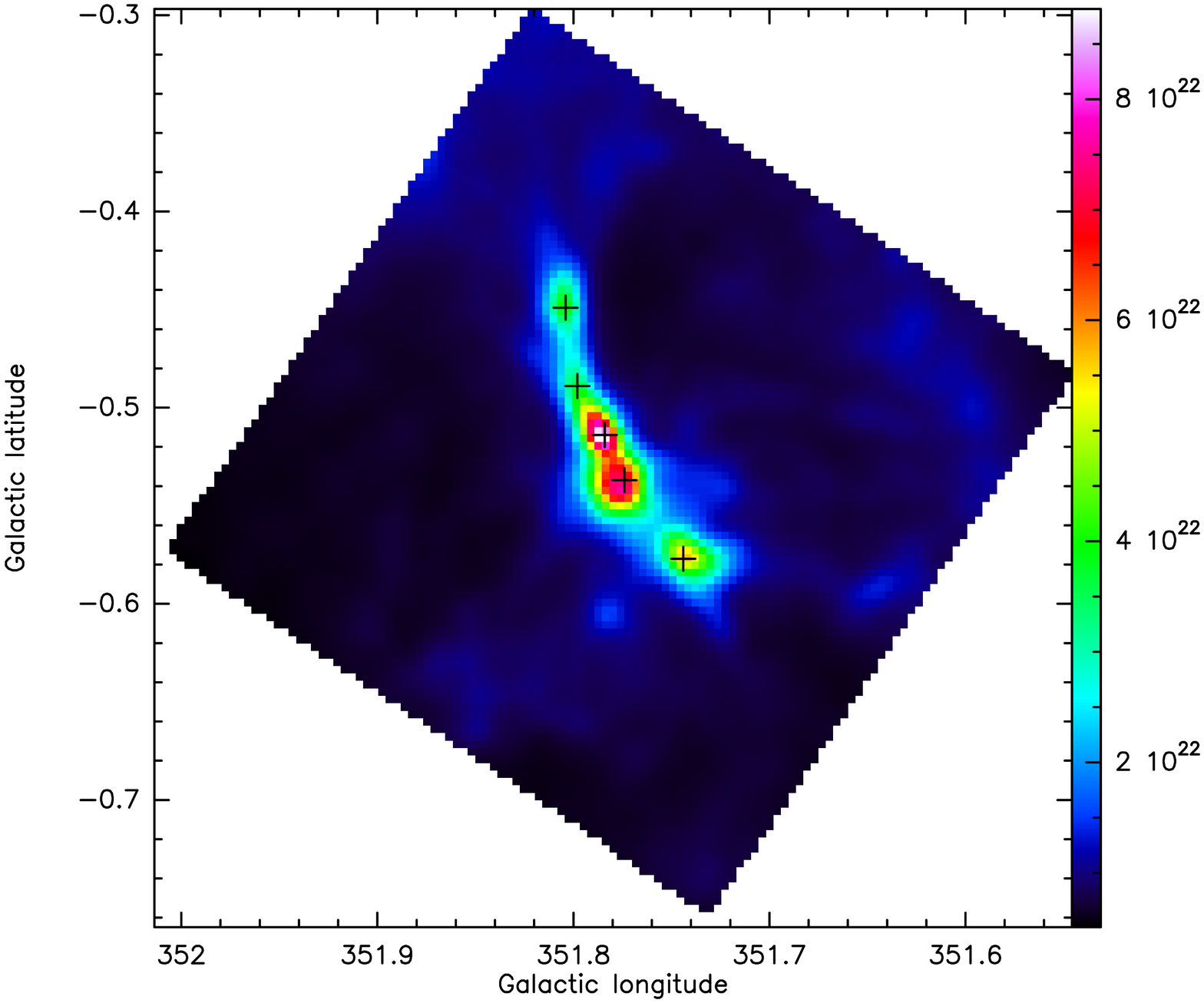,width=2.15in,height=1.5in}
\psfig{file=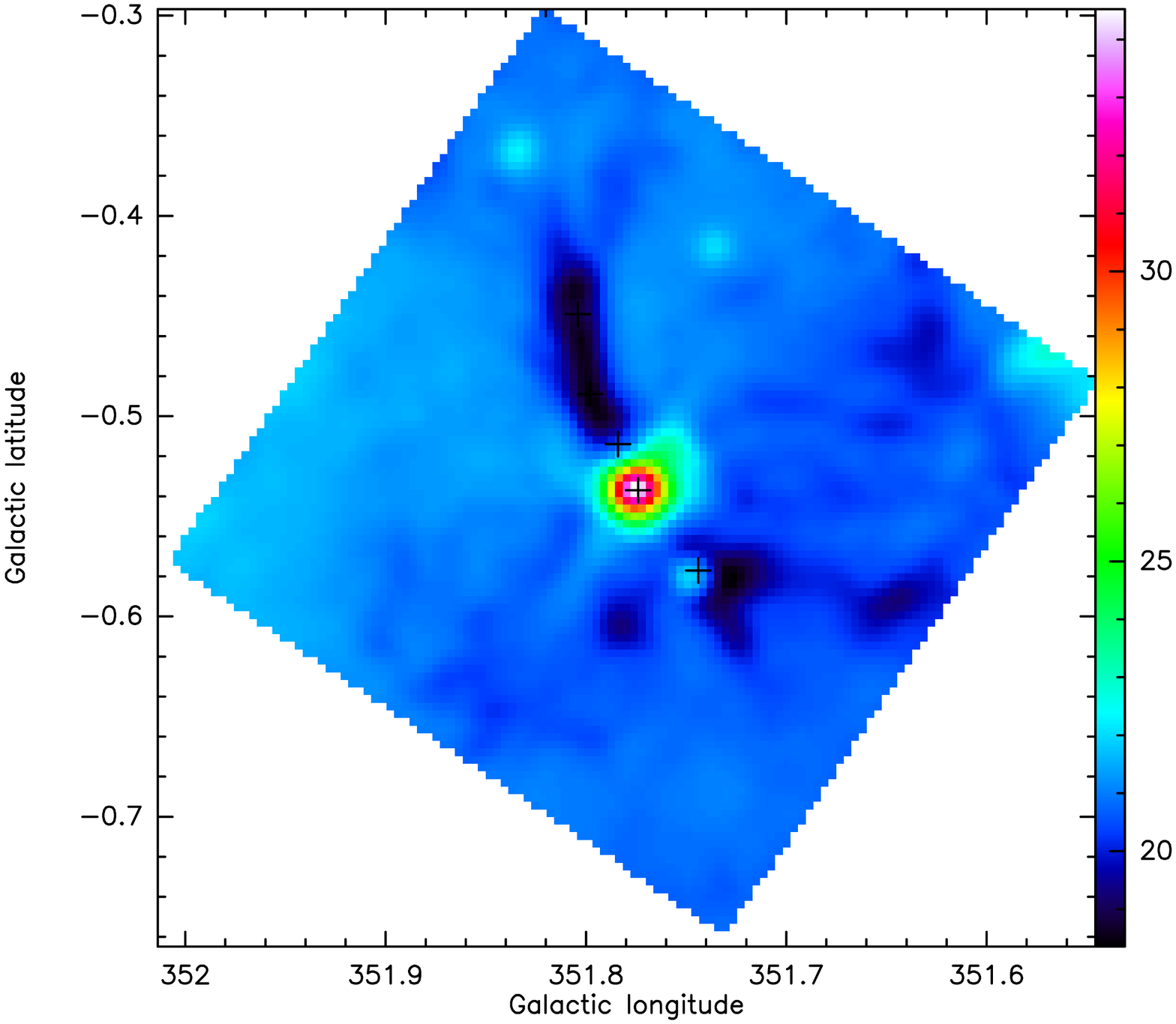,width=2.15in,height=1.5in}
\psfig{file=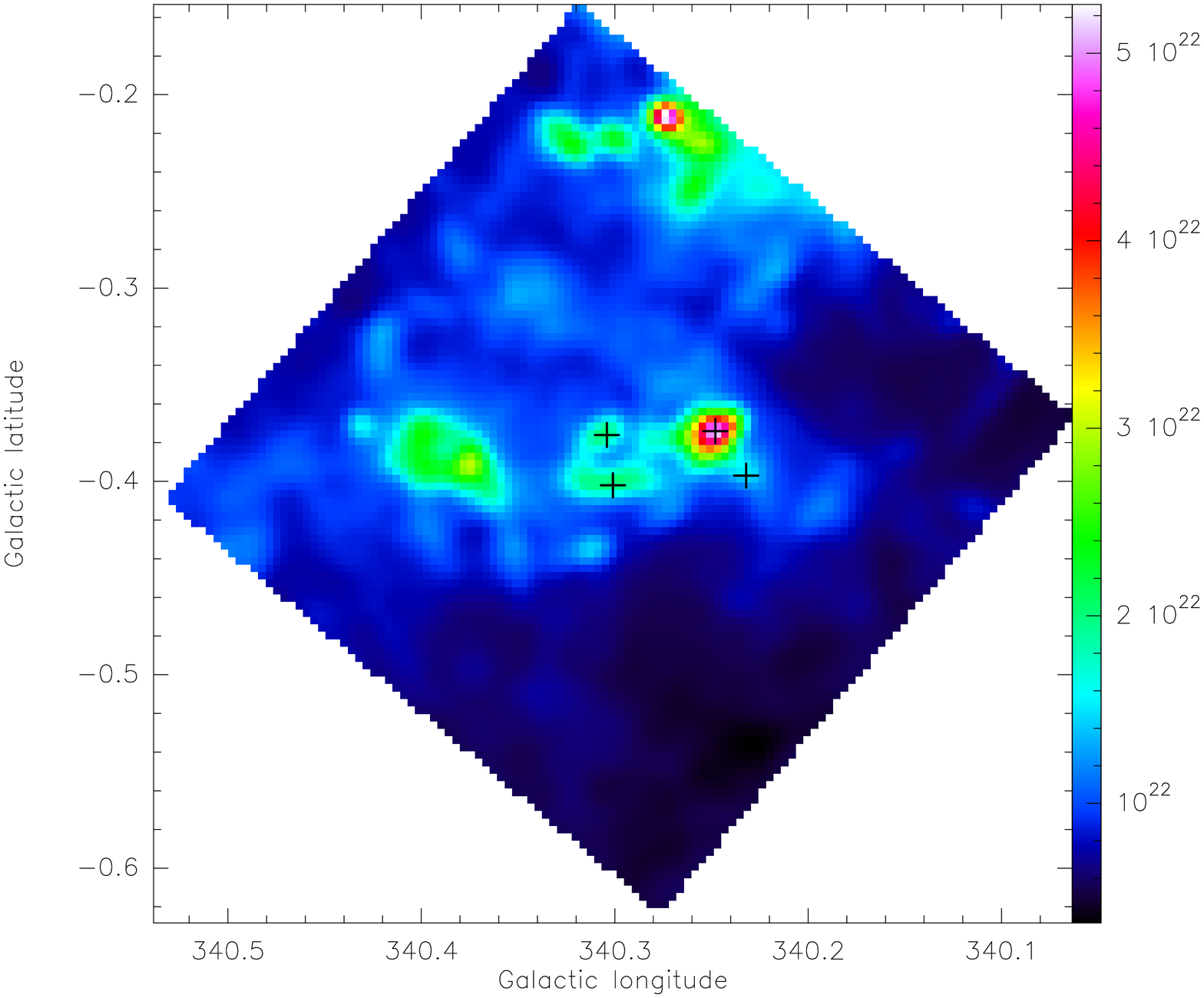,width=2.15in,height=1.5in}
\psfig{file=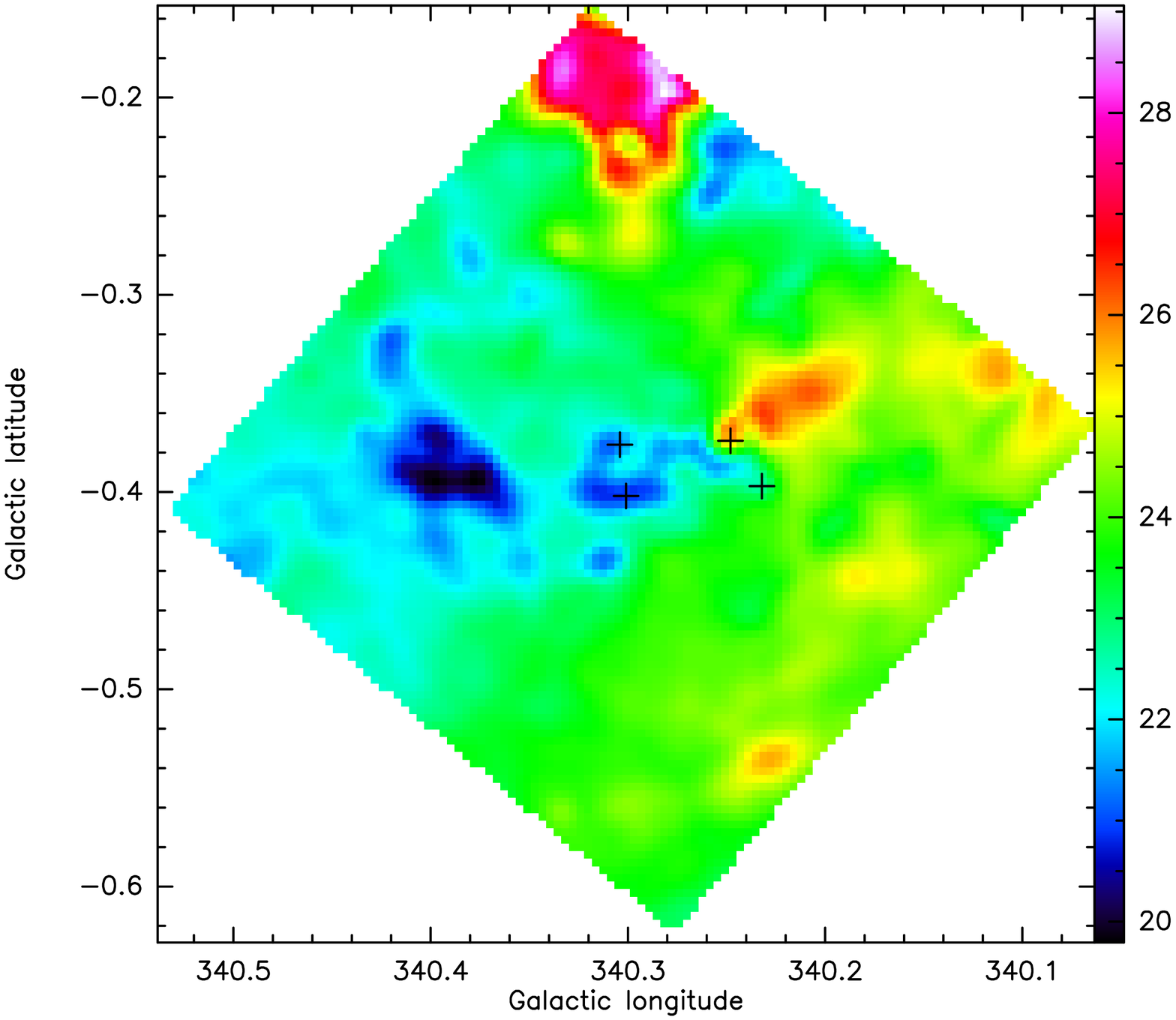,width=2.15in,height=1.5in}
\psfig{file=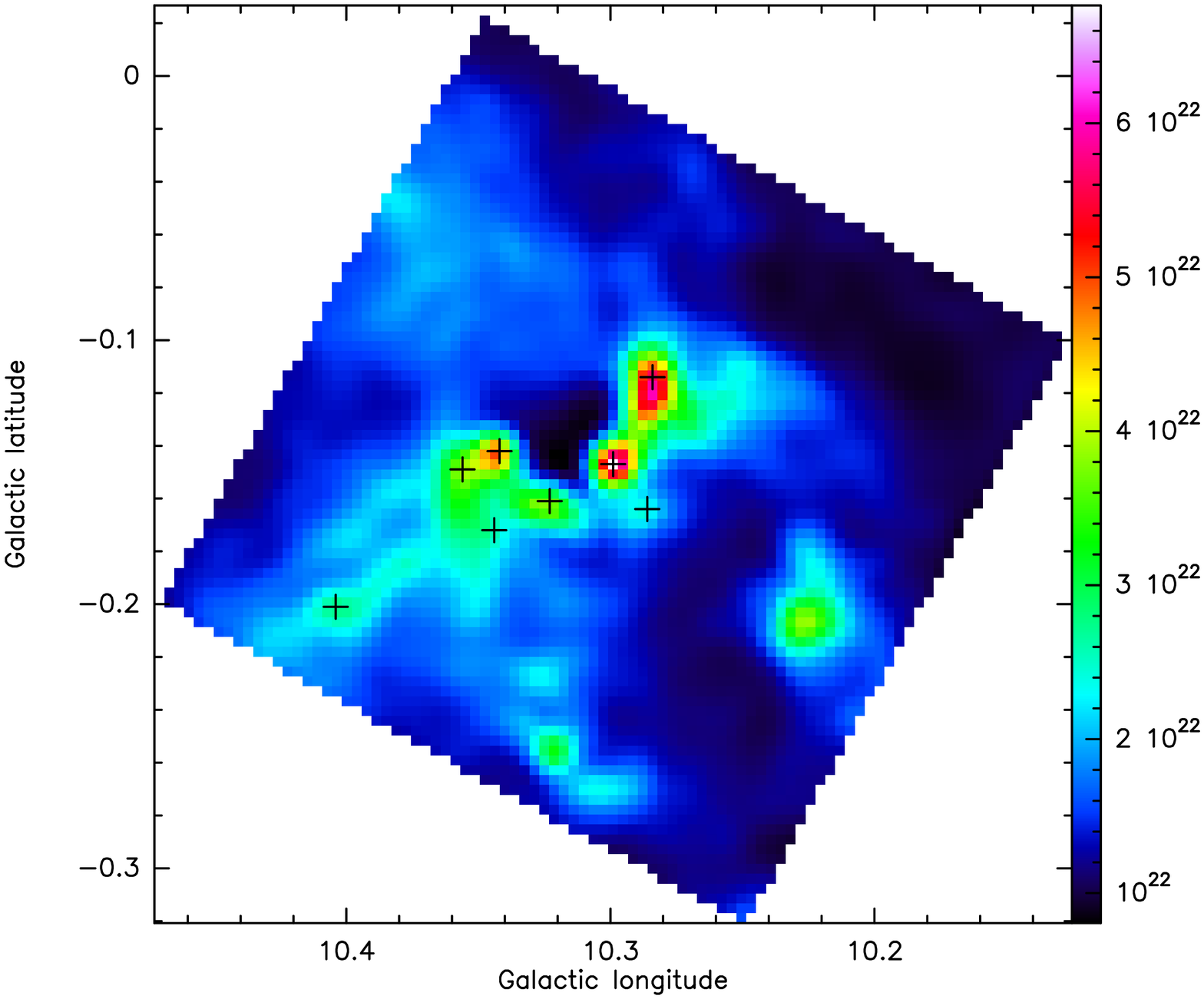,width=2.15in,height=1.5in}
\psfig{file=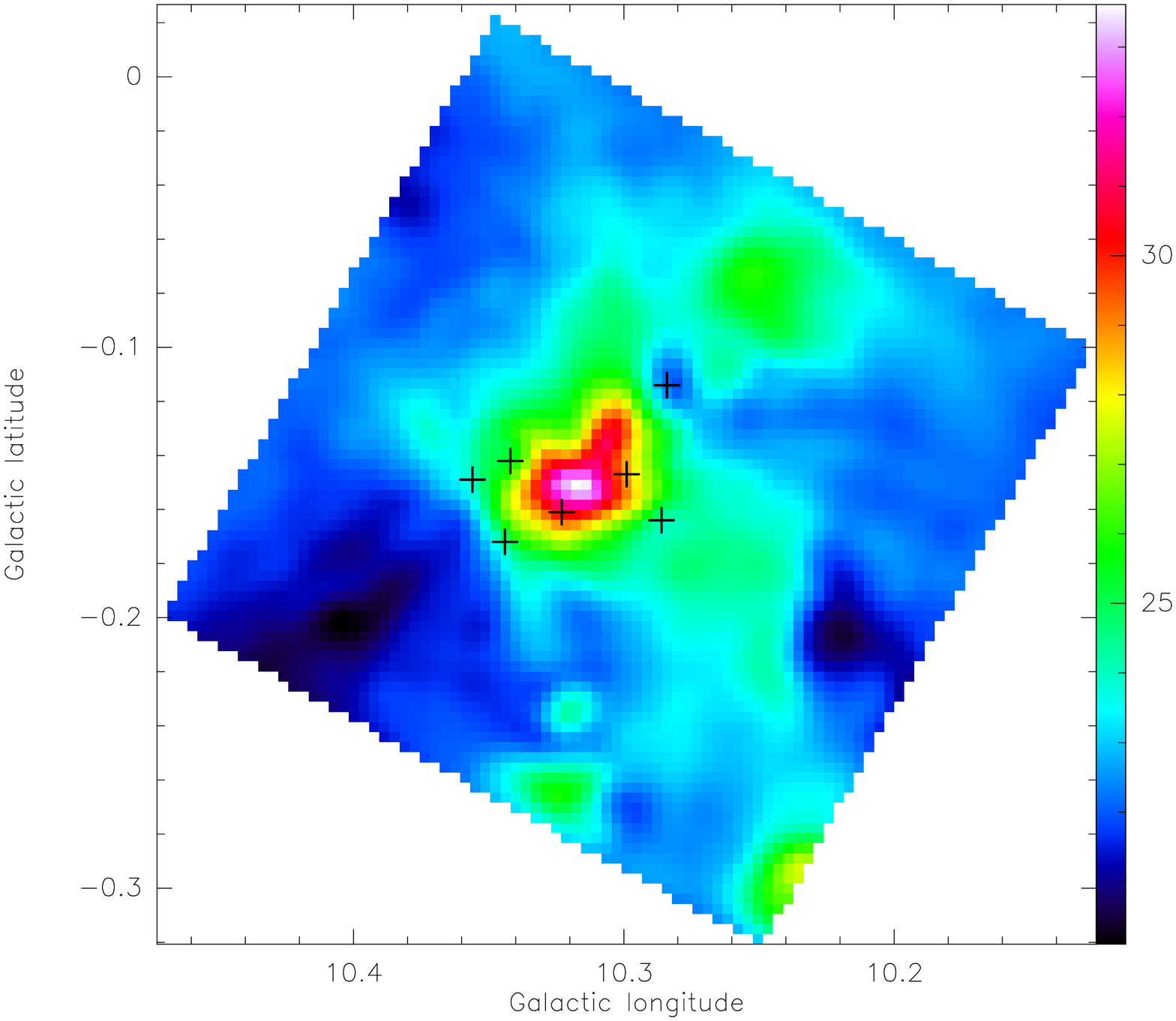,width=2.15in,height=1.5in}
\psfig{file=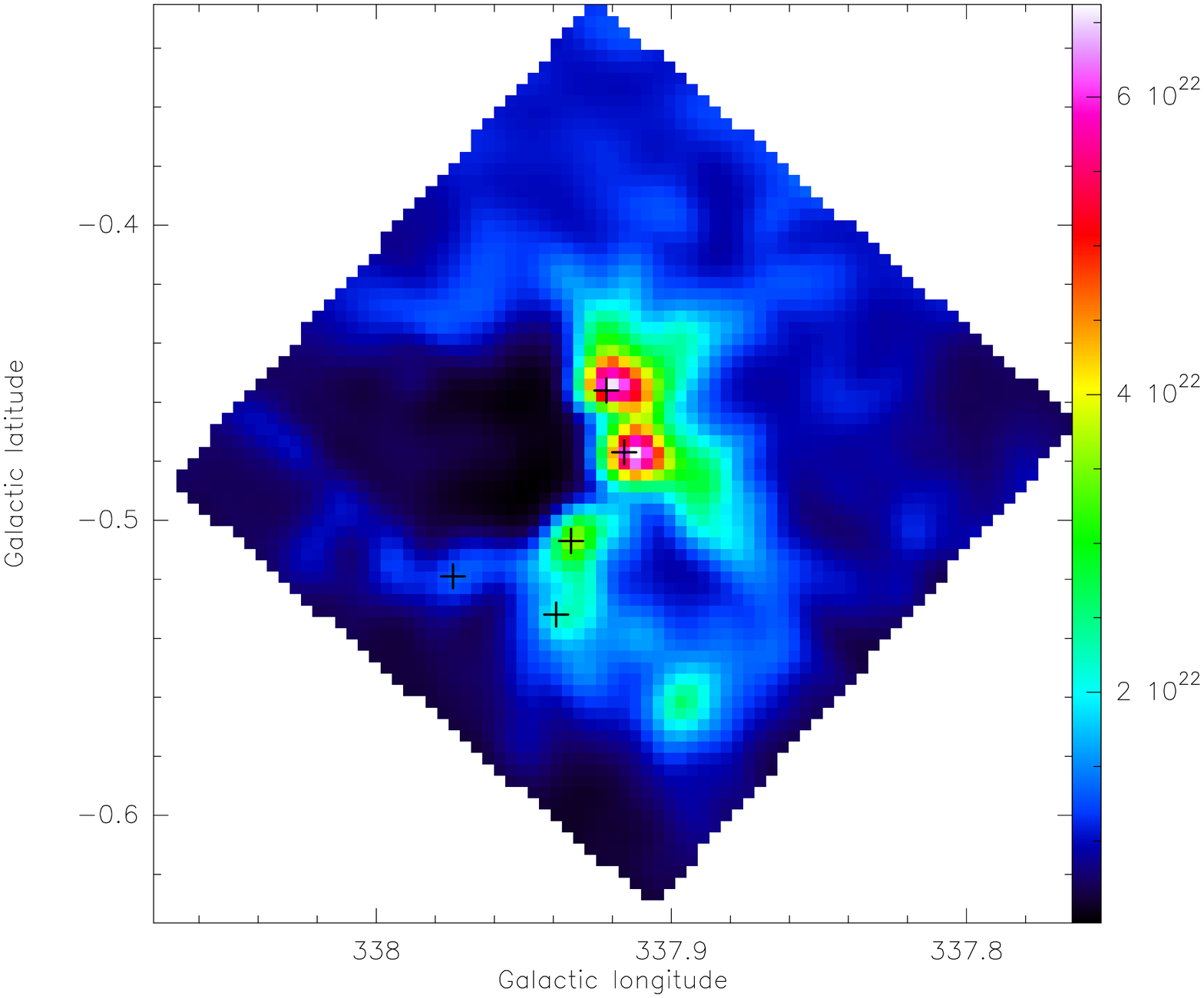,width=2.15in,height=1.5in}
\psfig{file=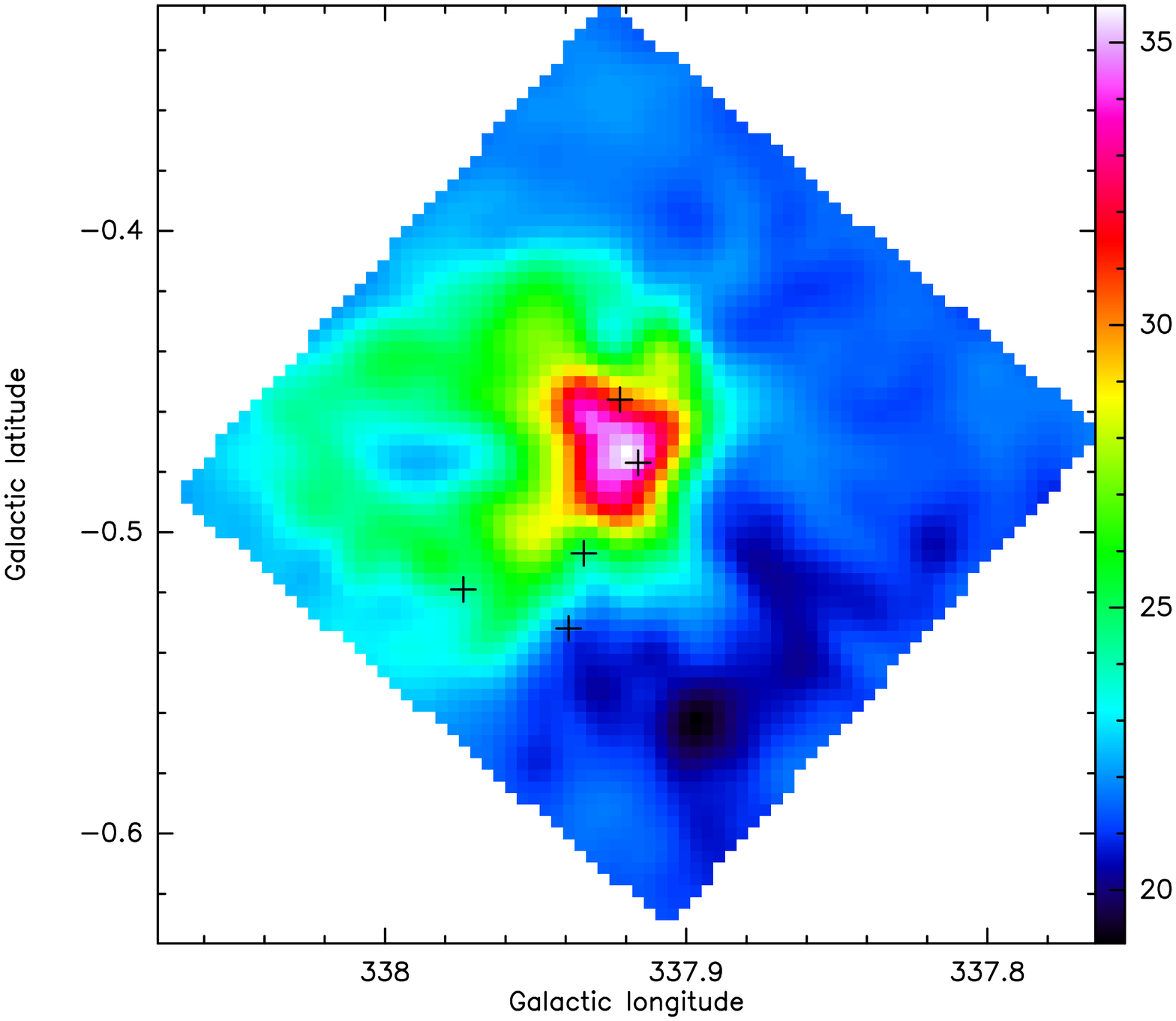,width=2.15in,height=1.5in}
\psfig{file=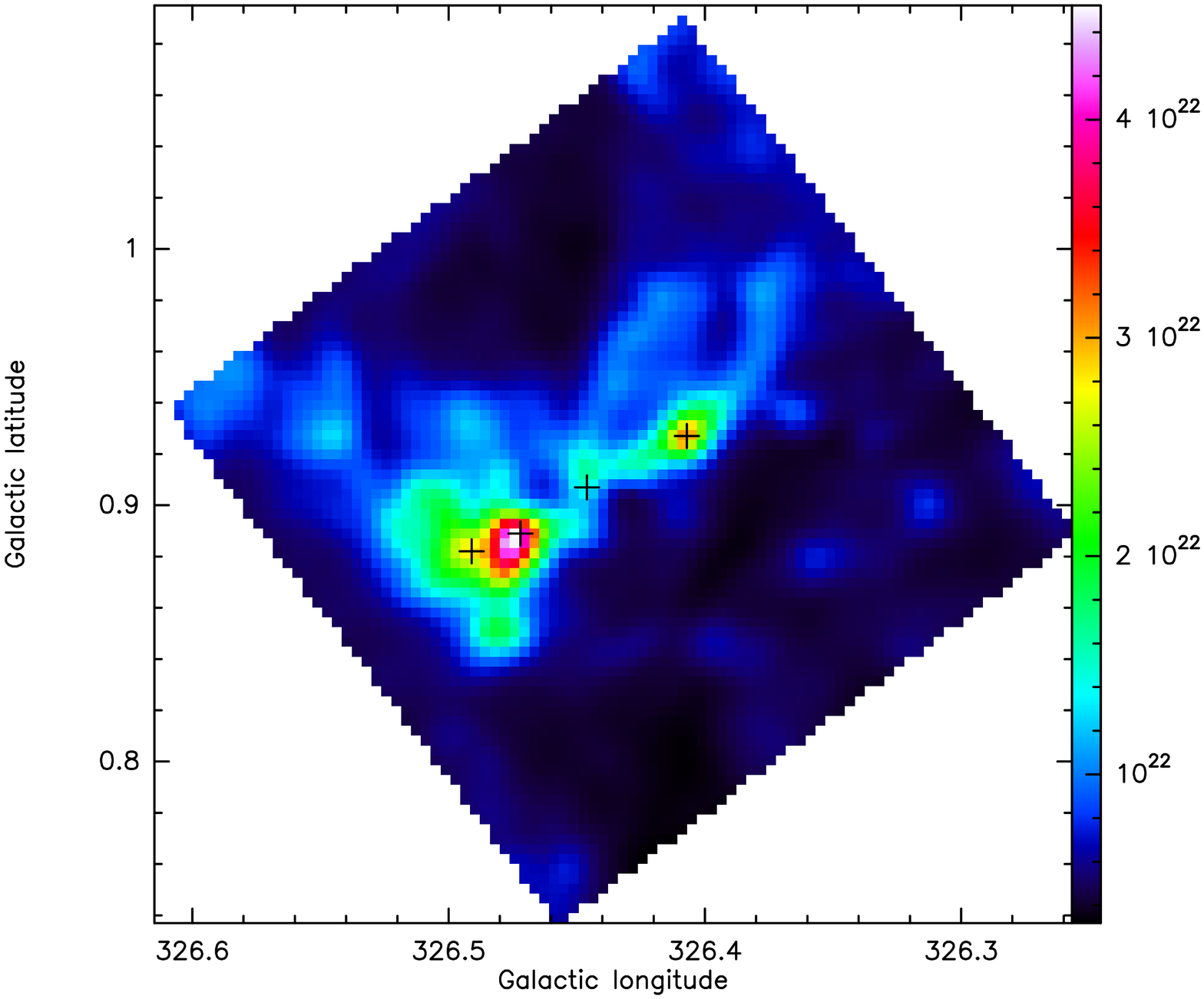,width=2.15in,height=1.5in}
\psfig{file=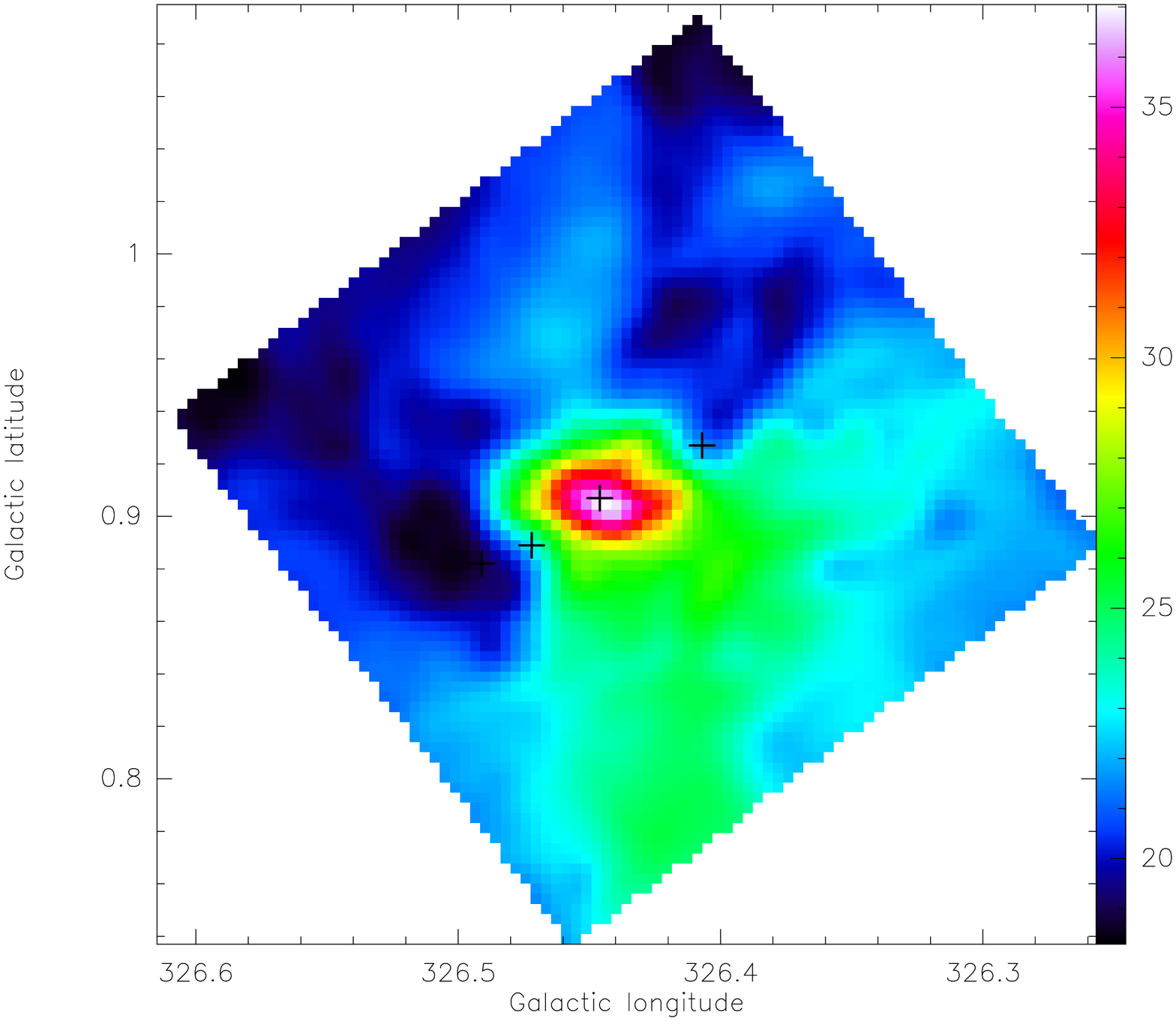,width=2.15in,height=1.5in}
\psfig{file=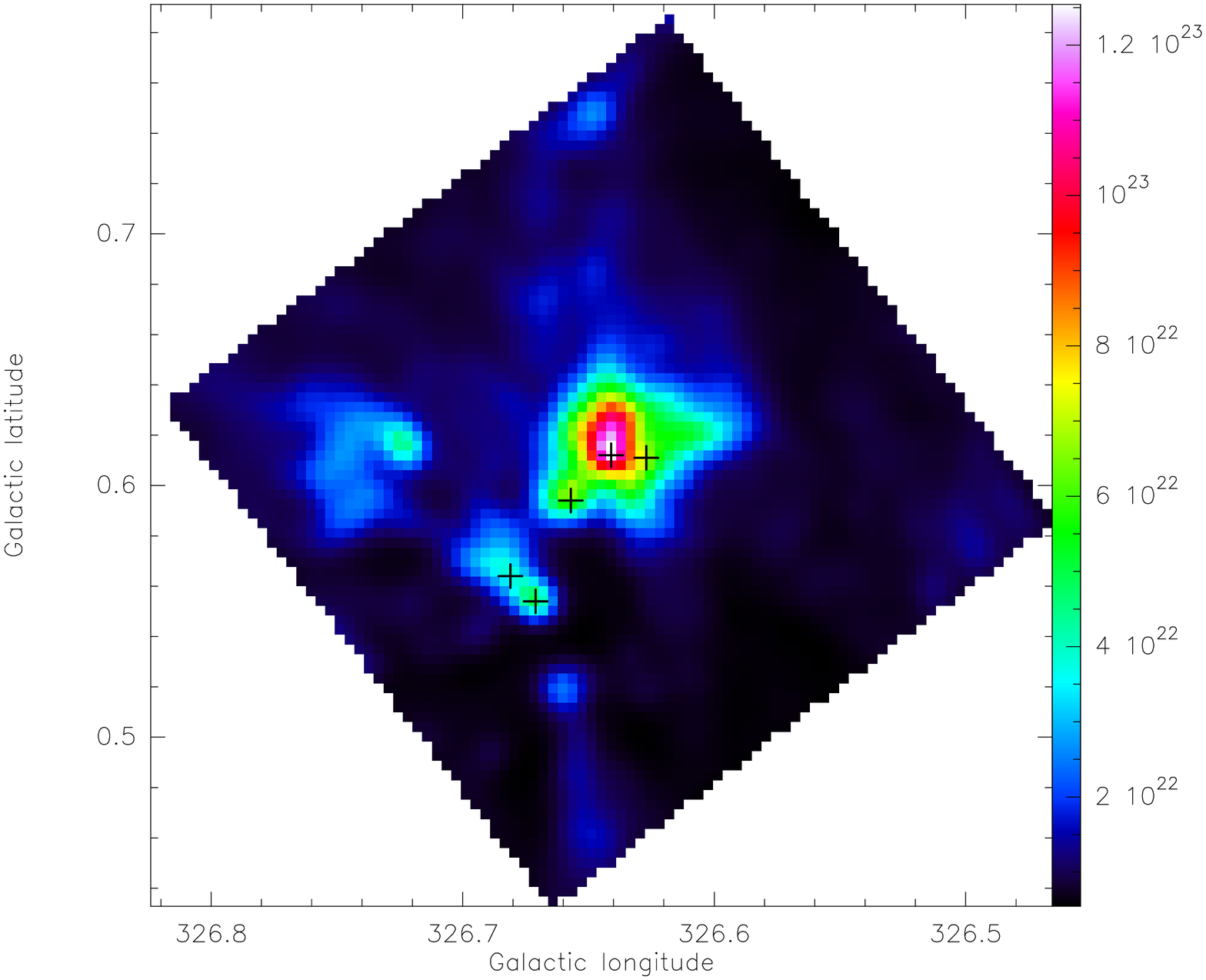,width=2.15in,height=1.5in}
\psfig{file=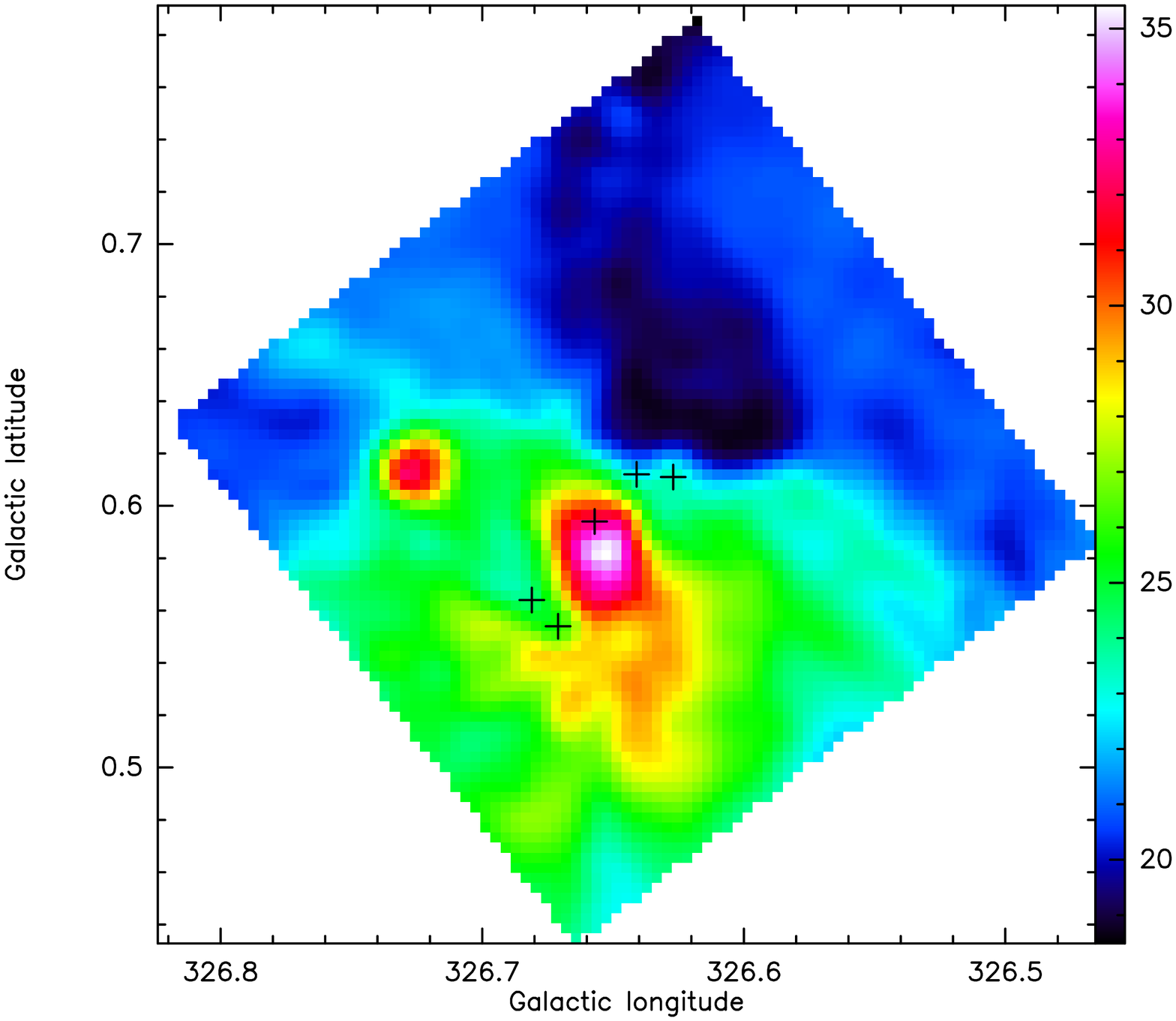,width=2.15in,height=1.5in} \caption{Left panels: H$_2$ column density of the six sources built on the SED fitting pixel by pixel. The pluses mark the dense clumps listed in table 1. The unit of each color bar is in cm$^{-2}$. Right panels: Dust temperature maps in color scale derived from SED fitting pixel by pixel. The unit of each color bar is in K.}
\end{figure}

%fig14
\begin{figure}
\center
\psfig{file=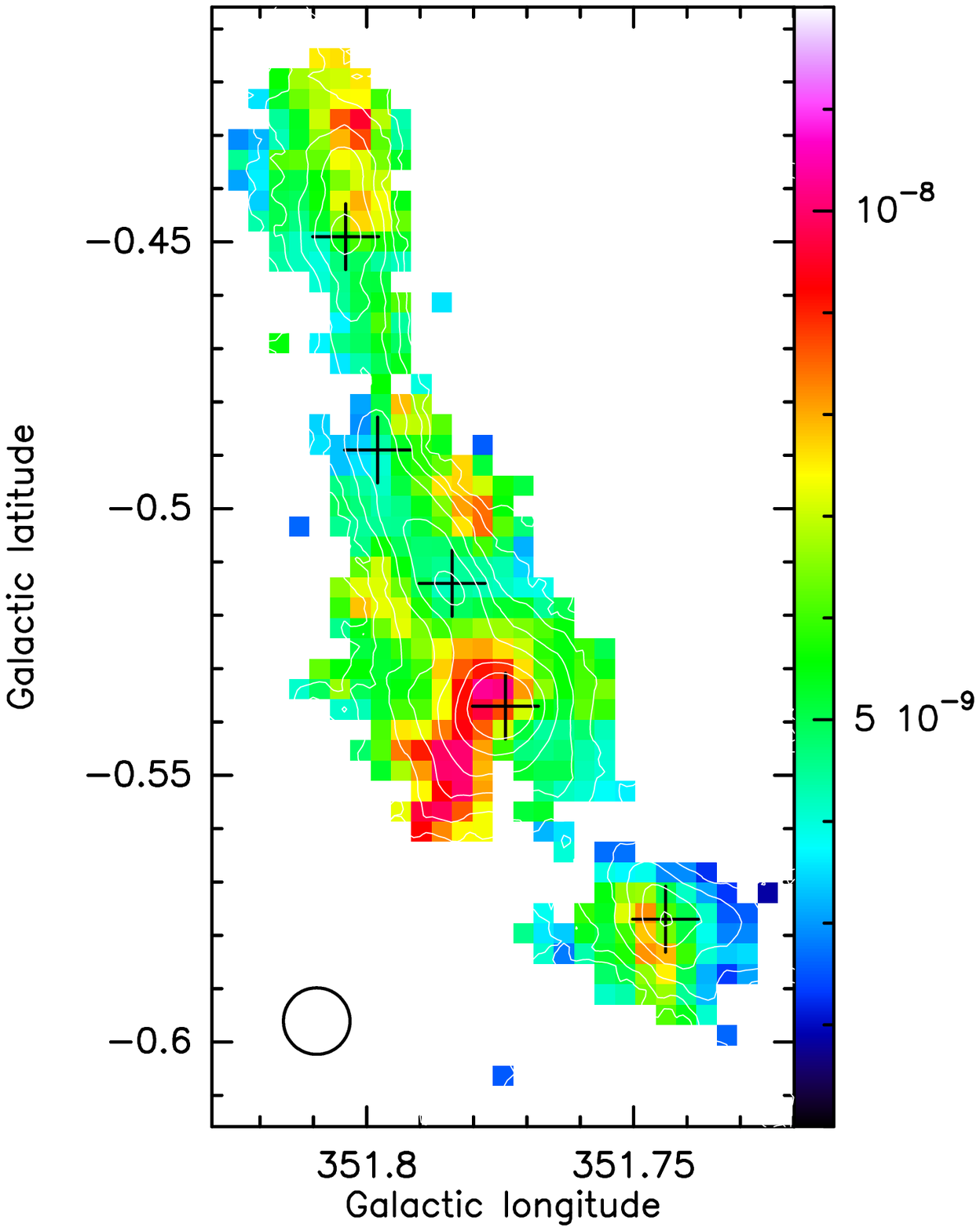,width=2.3in,height=2.9in}
\psfig{file=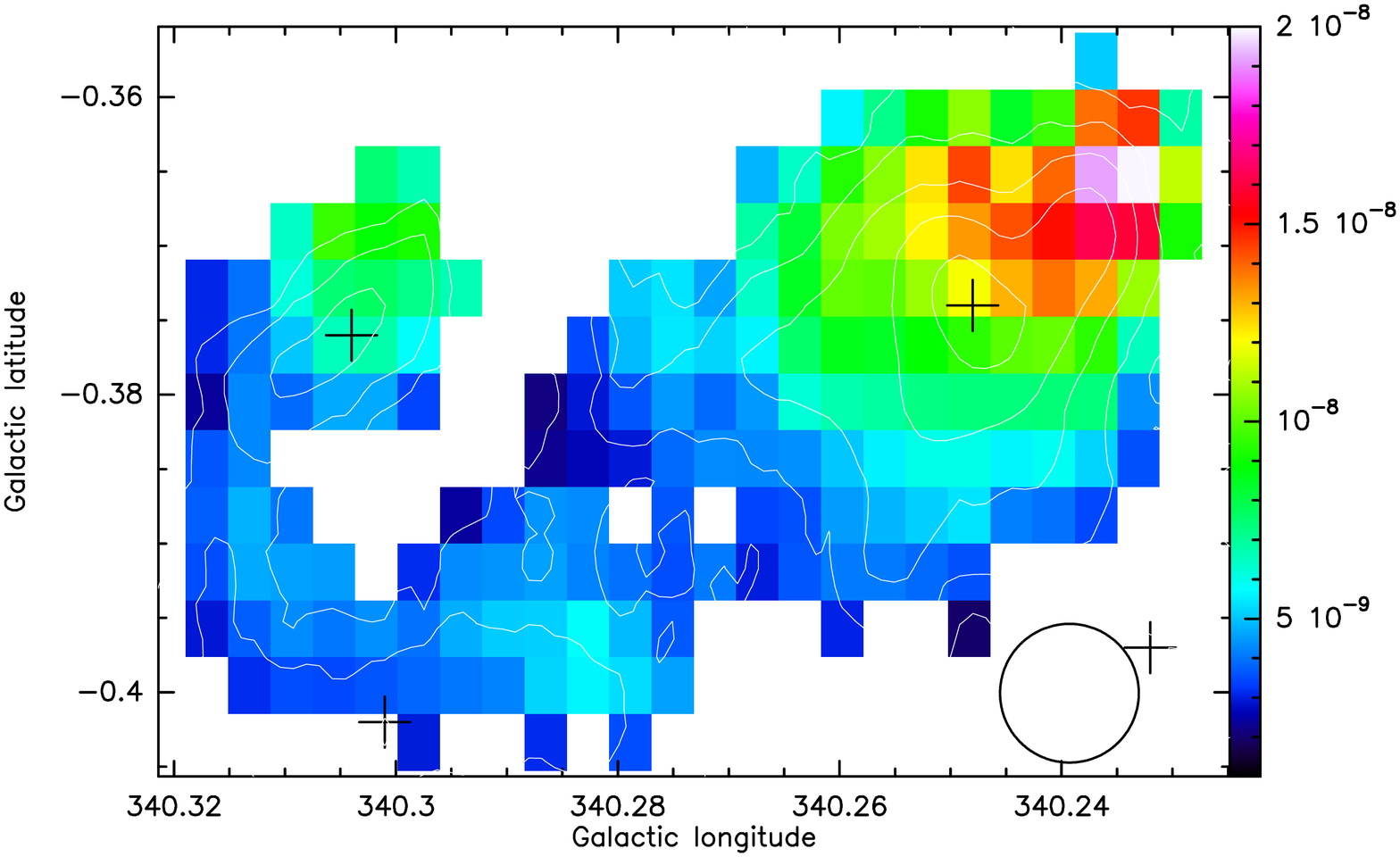,width=2.9in,height=1.8in}
\psfig{file=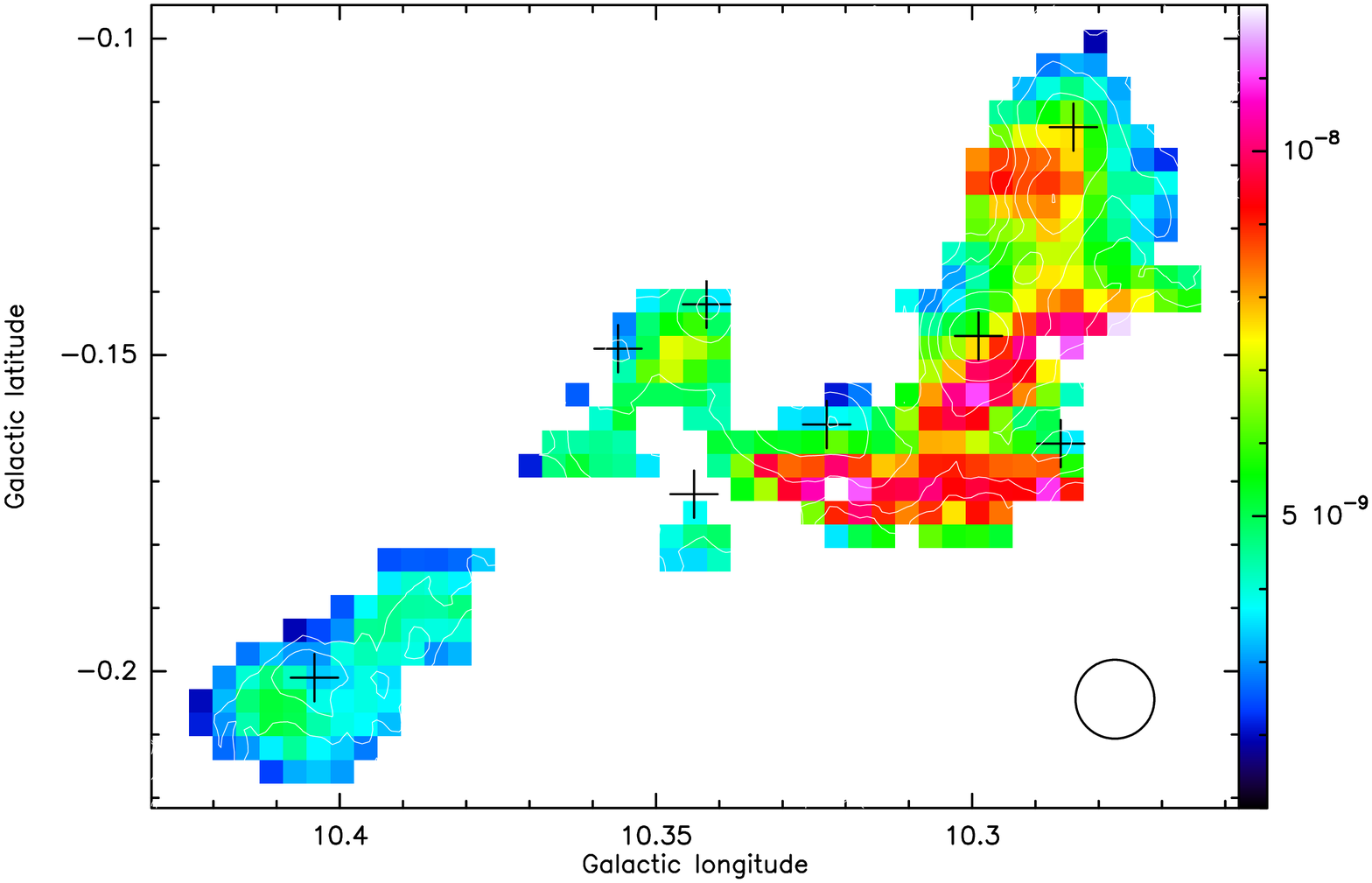,width=3in,height=2in}
\psfig{file=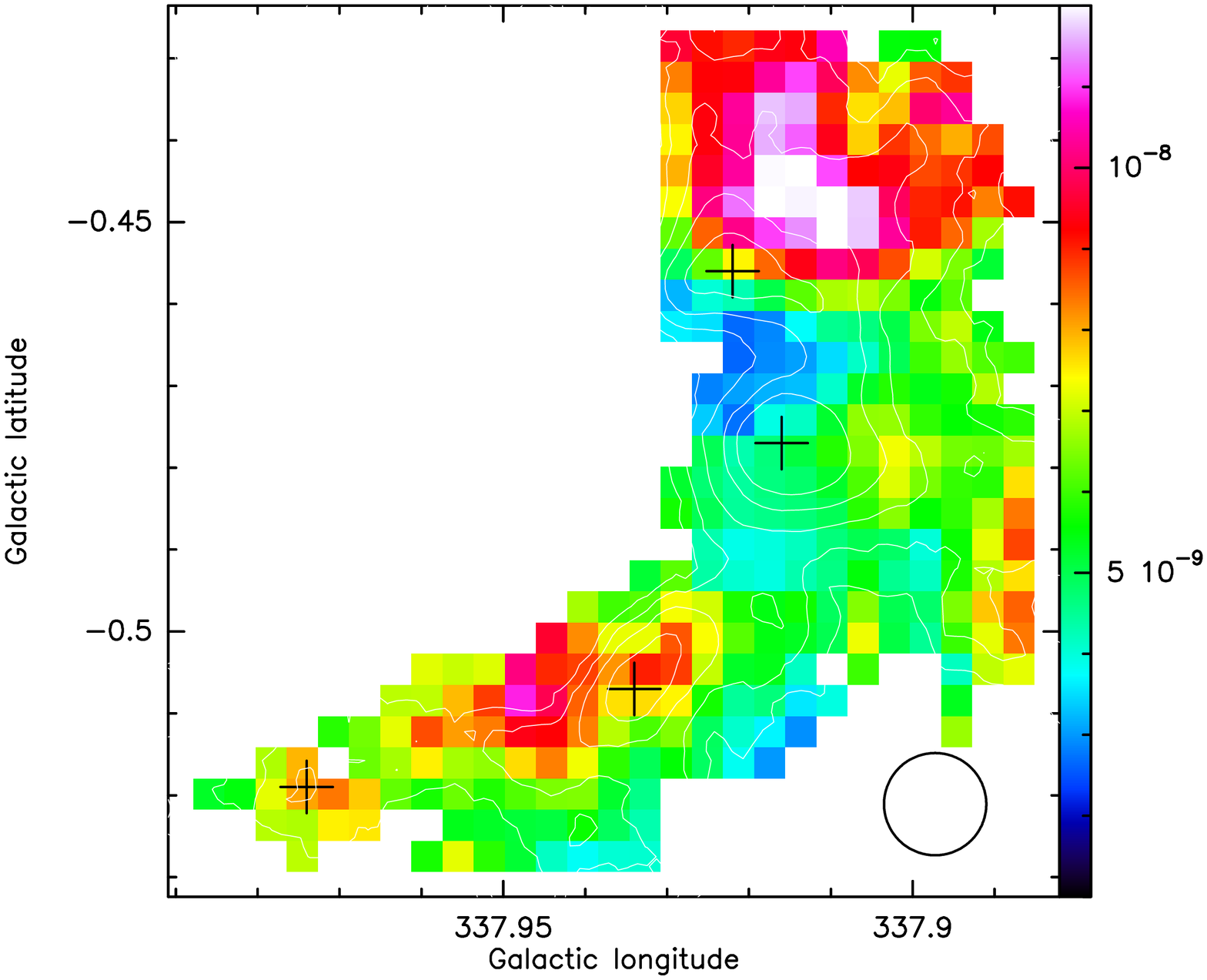,width=2.4in,height=2in}
\psfig{file=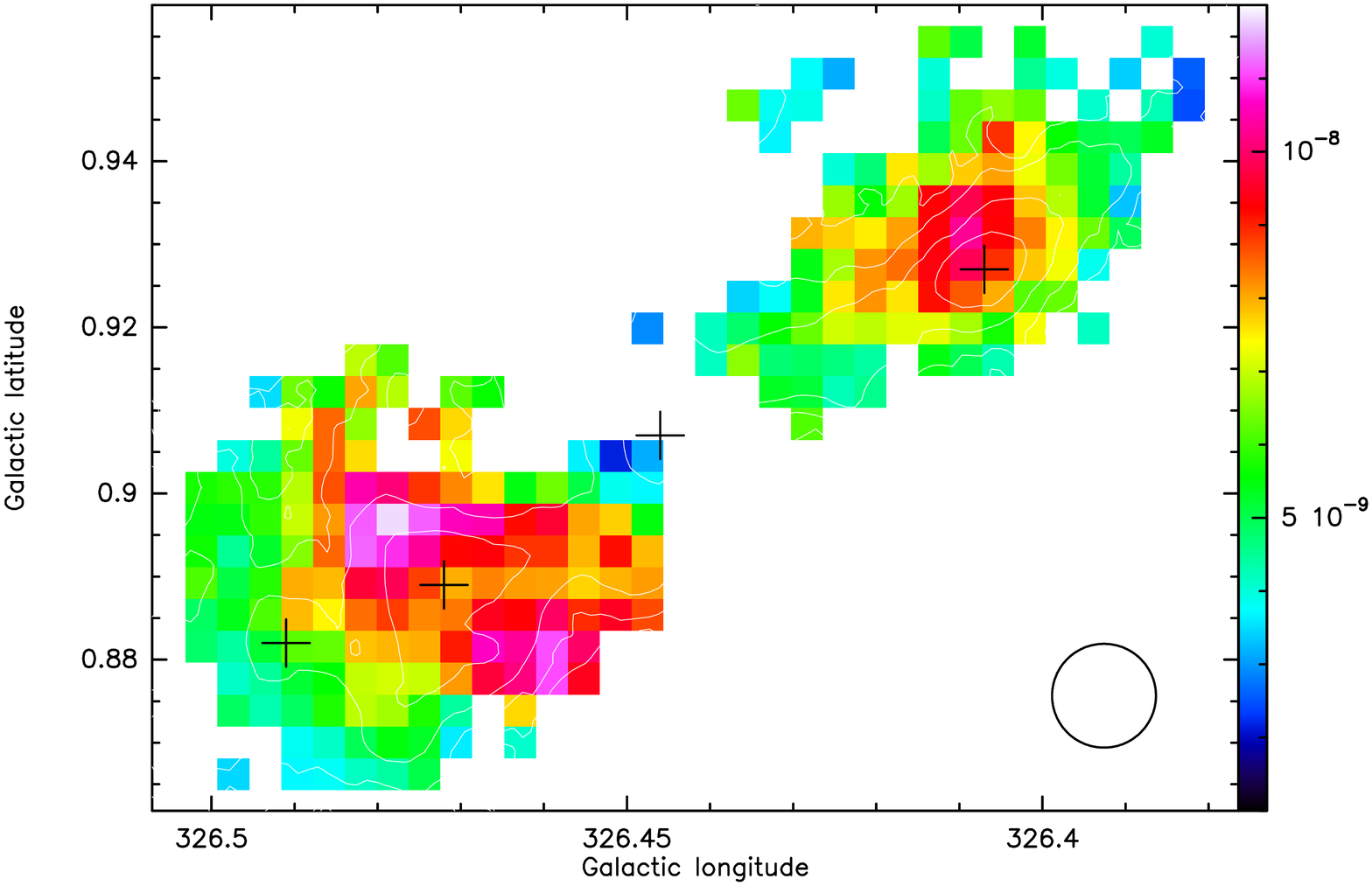,width=3in,height=2in}
\psfig{file=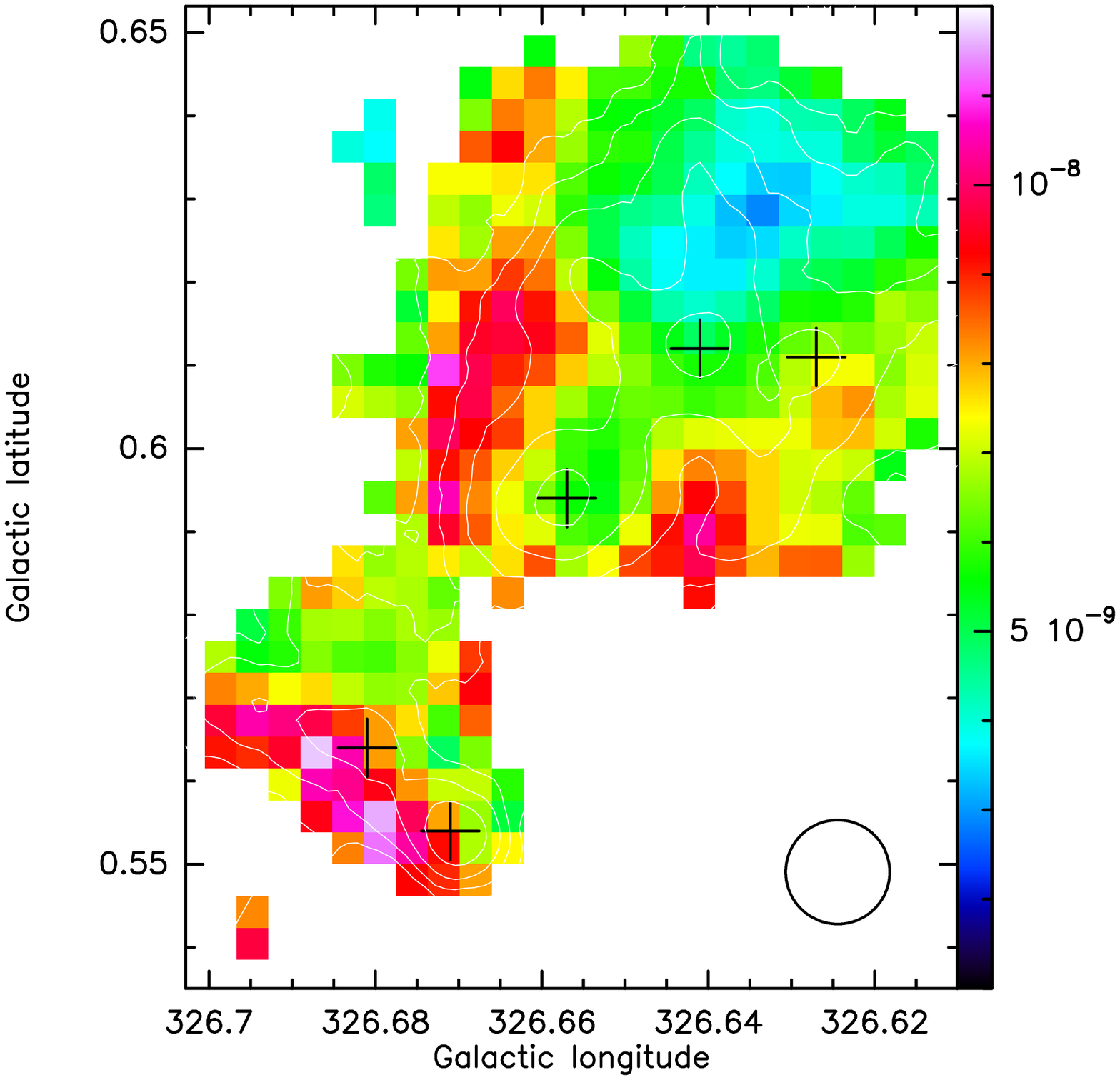,width=2.1in,height=2in} \caption{Calculated N$_2$H$^+$ abundance maps of the six sources. The pluses mark the dense clumps listed in table 1. The black circle shown in each image indicates the beam size of 45$^\prime$$^\prime$. }
\end{figure}

%fig15
\begin{figure}
\center
\psfig{file=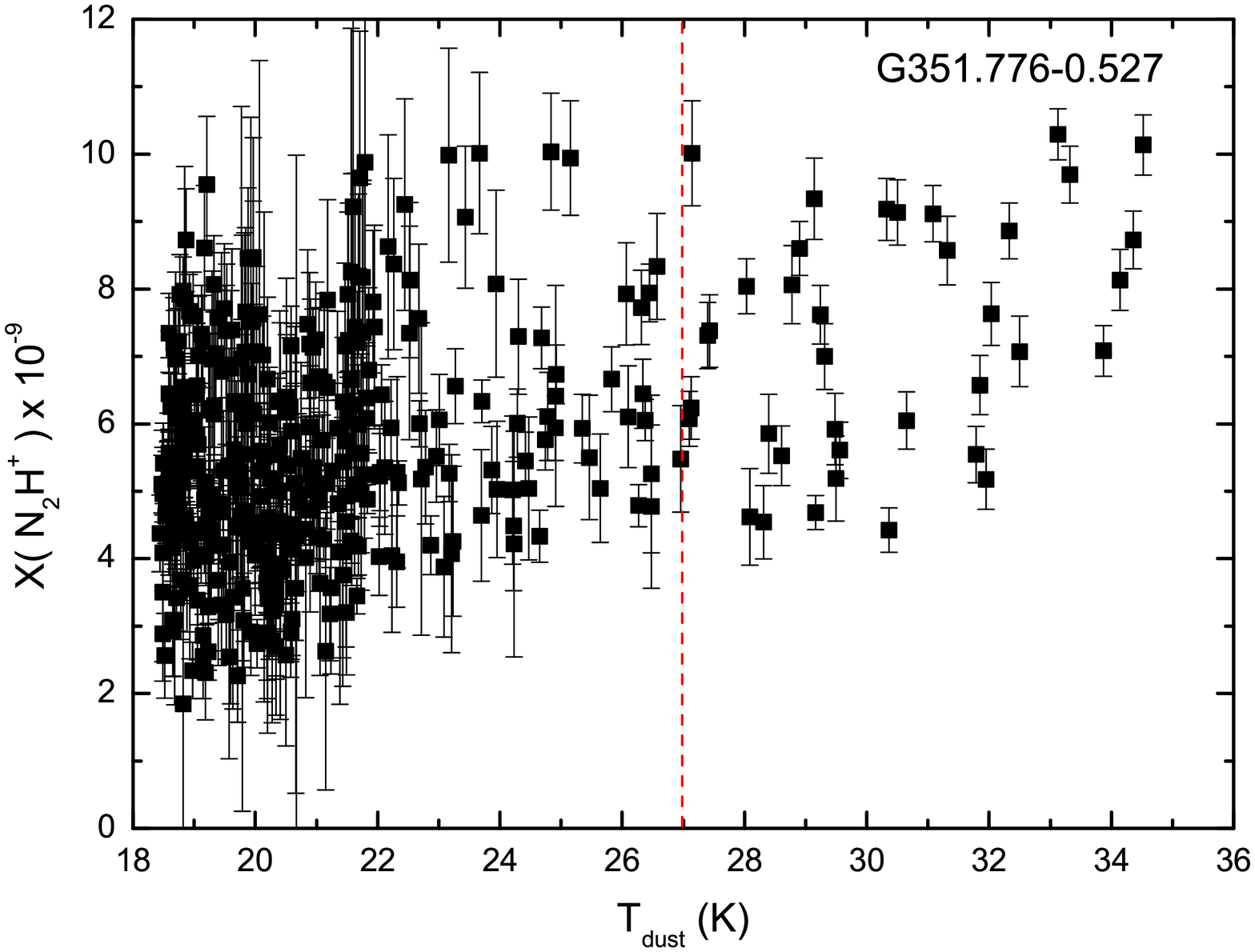,width=3in,height=2in}
\psfig{file=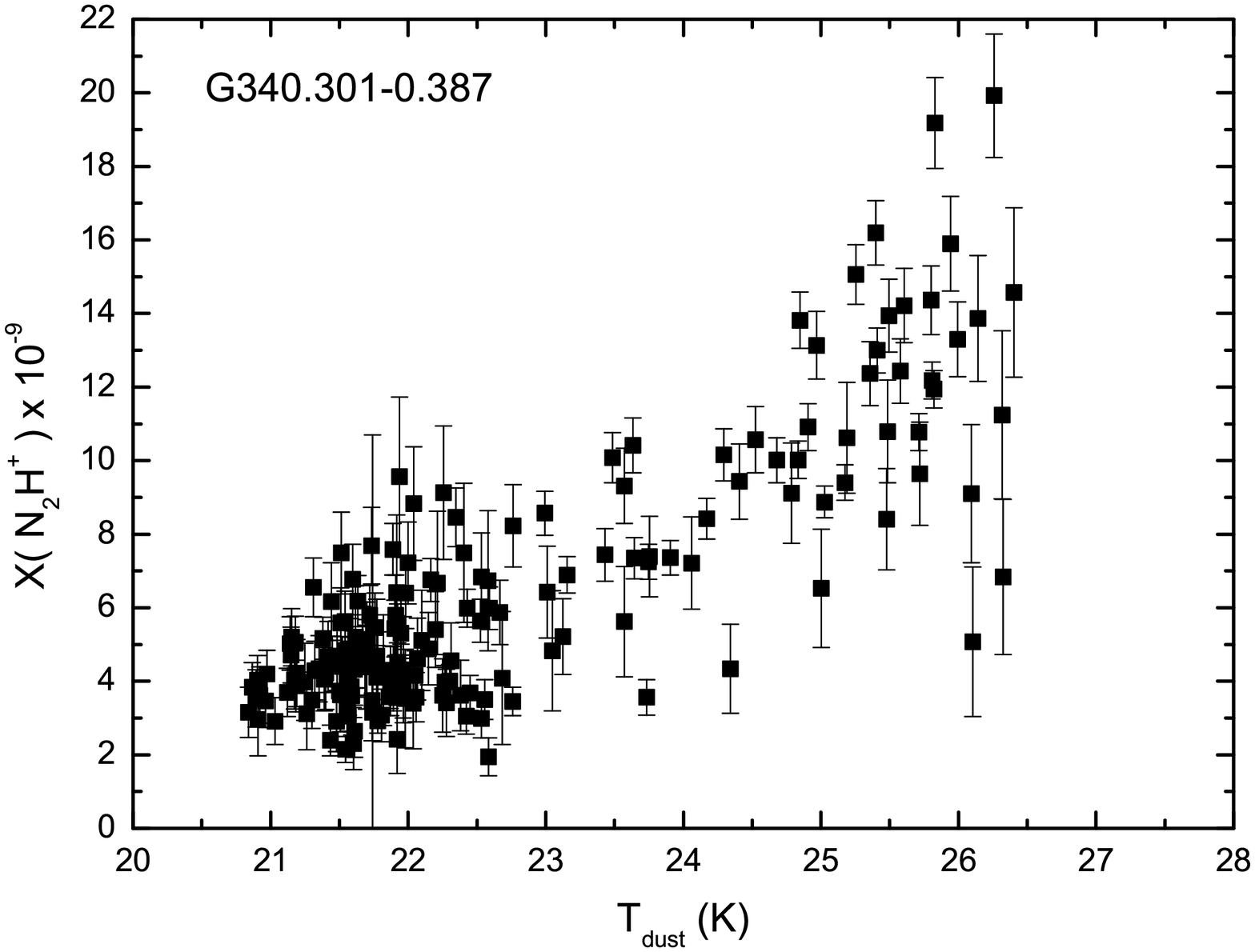,width=3in,height=2in}
\psfig{file=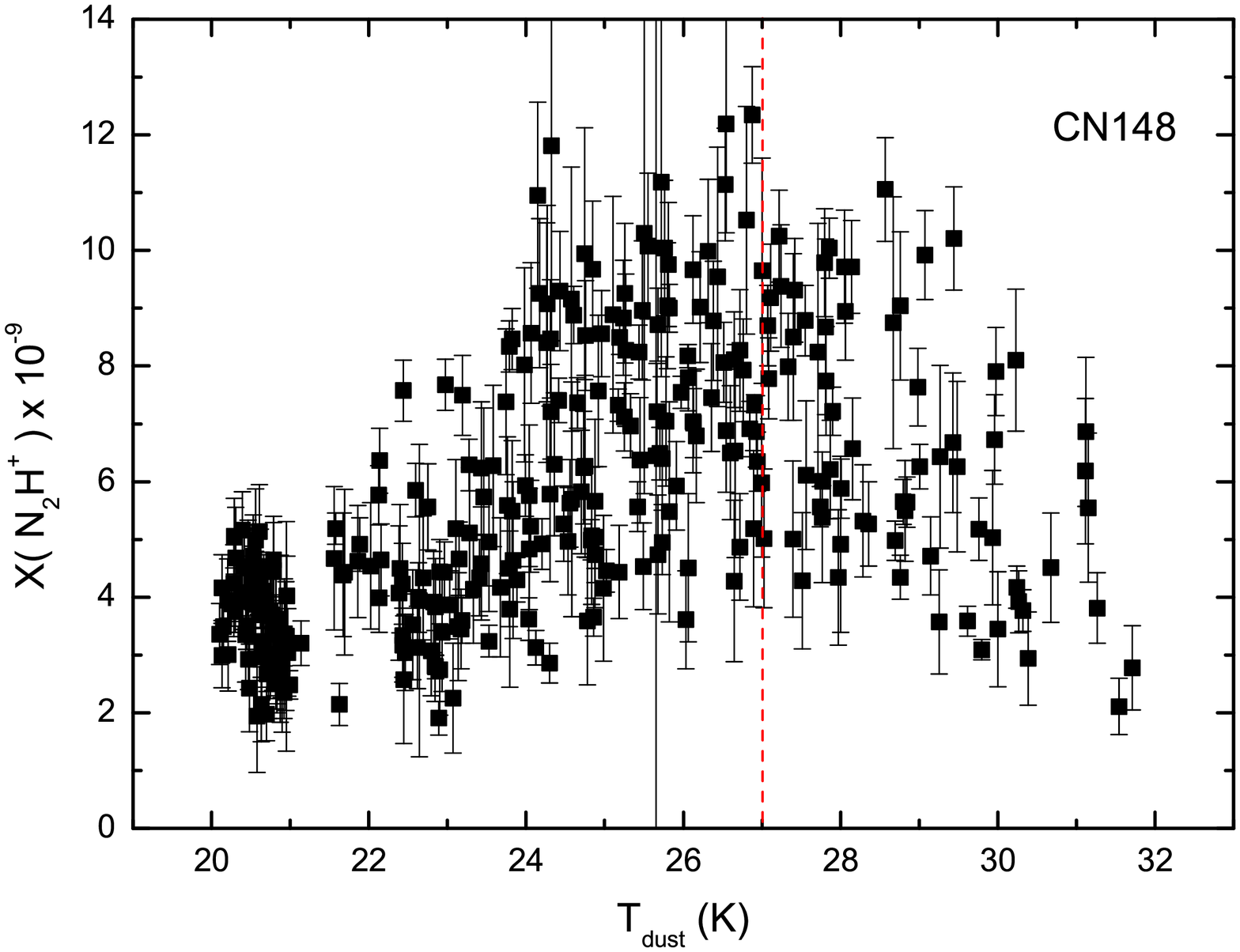,width=3in,height=2in}
\psfig{file=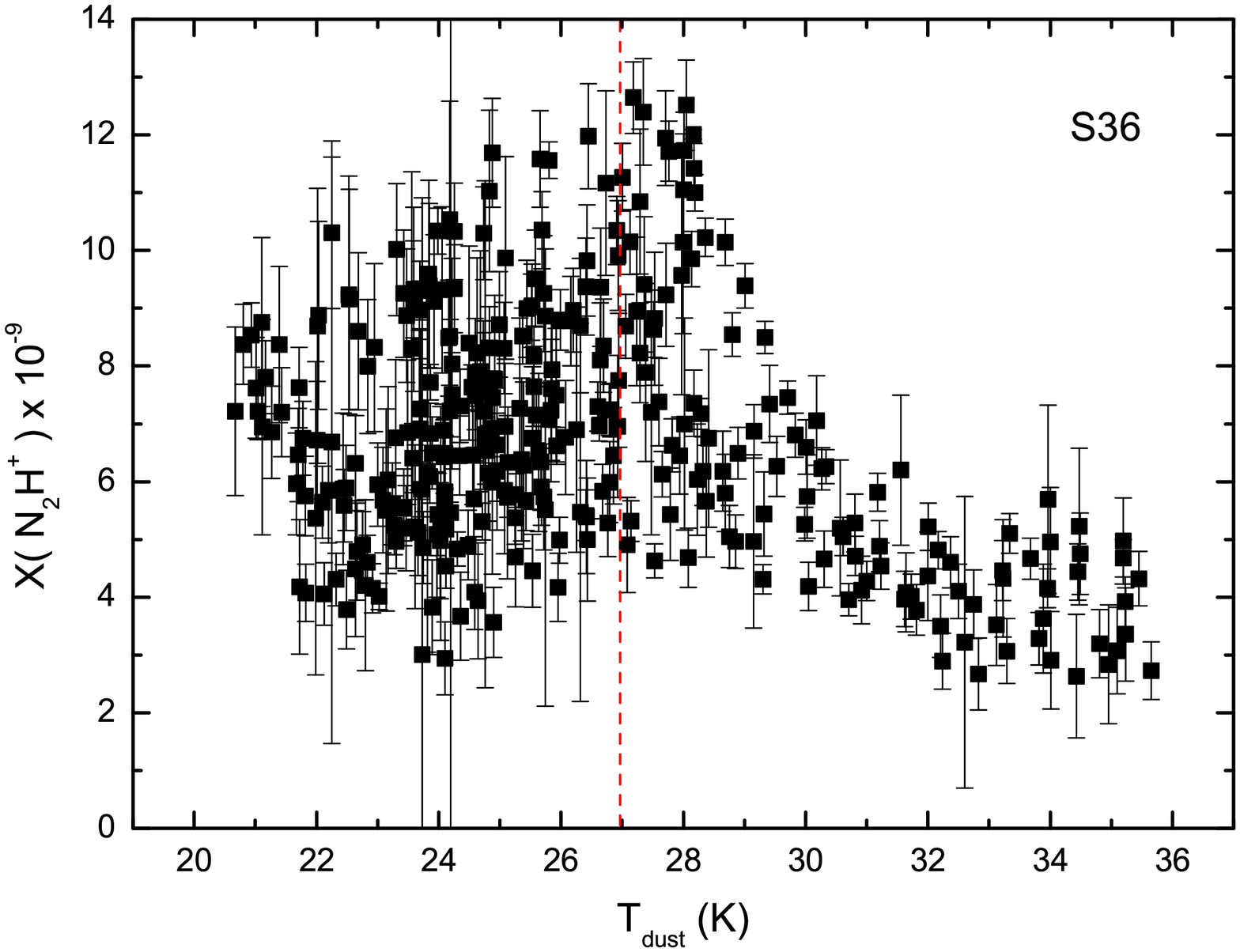,width=3in,height=2in}
\psfig{file=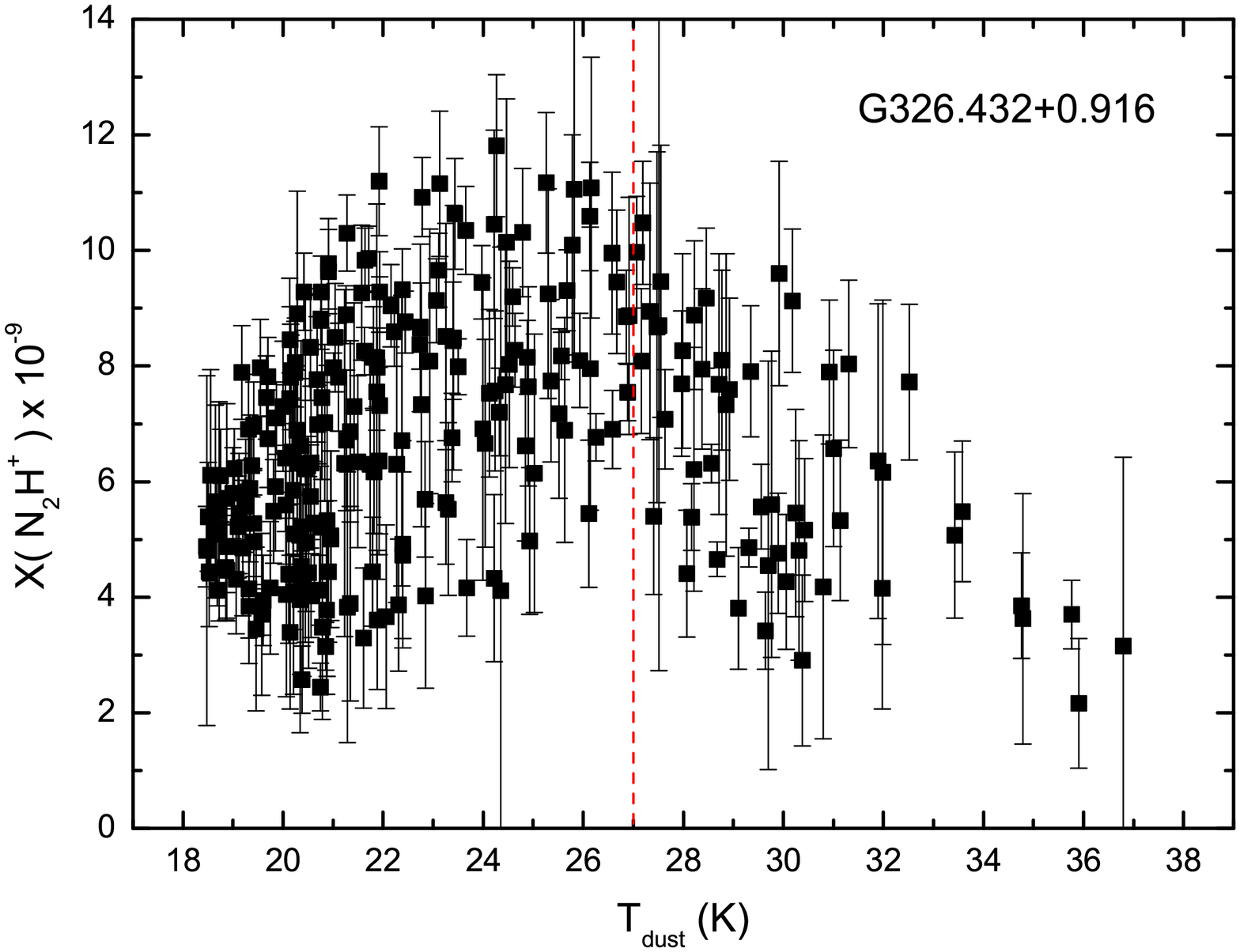,width=3in,height=2in}
\psfig{file=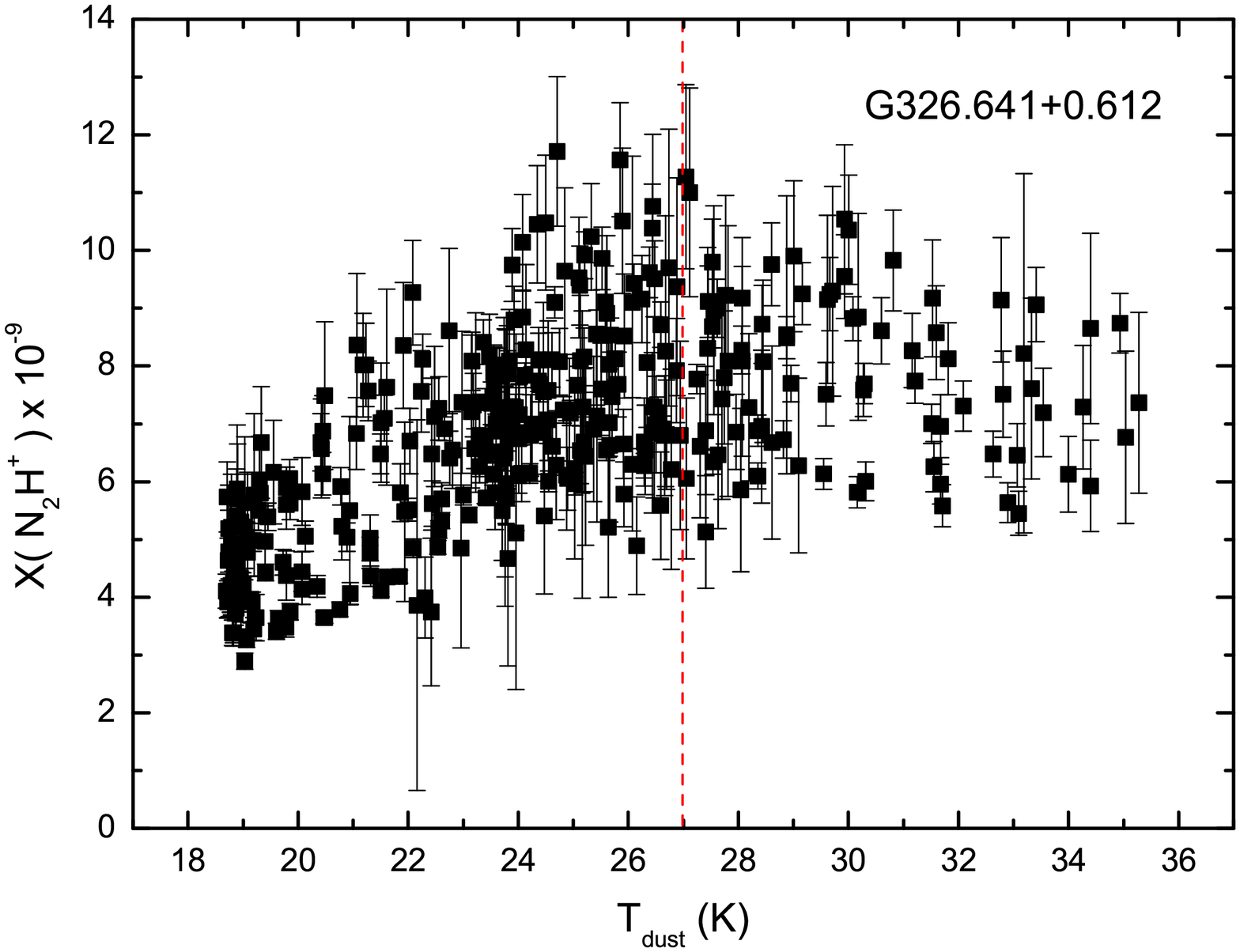,width=3in,height=2in} \caption{Abundance of N$_2$H$^+$ plotted as a function of dust temperature in each pixel of the six sources. The red dashed lines mark T$_d$ = 27 K. }
\end{figure}

\end{document}